\documentclass[pdflatex,sn-nature]{sn-jnl}

\usepackage{graphicx}%
\usepackage{multirow}%
\usepackage{amsmath,amssymb,amsfonts}%
\usepackage{amsthm}%
\usepackage{mathrsfs}%
\usepackage[title]{appendix}%
\usepackage{xcolor}%
\usepackage{textcomp}%
\usepackage{manyfoot}%
\usepackage{booktabs}%
\usepackage{algorithm}%
\usepackage{algorithmicx}%
\usepackage{algpseudocode}%
\usepackage{listings}%

\begin{document}
\title{Density fluctuations for Squeezed Number State and Coherent Squeezed Number State in Flat FRW Universe}
\author[1]{\fnm{Dhwani} \sur{Gangal}}

\author[1]{\fnm{Sudhava} \sur{Yadav}}

\author*[1]{\fnm{K.K.} \sur{Venkataratnam}}\email{kvkamma.phy@mnit.ac.in}

\affil[1]{\orgdiv{Department of Physics}, \orgname{Malaviya National Institute of Technology Jaipur}, \orgaddress{\street{Jaipur}, \city{Jaipur}, \postcode{302017}, \country{India}}}

\abstract{We study the density fluctuations for Coherent Squeezed Number State (CSNS) and Squeezed Number States (SNS) formalism in Semiclassical theory of gravity in flat FRW universe. We used Number state evolution of oscillatory phase of inflaton for coherent squeezed number state and squeezed number states formalisms. We analyzed that density fluctuations for SNS depends upon squeezing parameter and number state while for CSNS density fluctuations depends upon squeezing parameter, number state and coherent state parameter. These parameters plays an important role for quantum consideration of SNS and CSNS. The results of the analysis shows that increase in density fluctuations for both SNS and CSNS, demonstrate quantum behavior of SCEE as well as production of various kind of particles in these states.}

\maketitle

\section{Introduction}\label{S1}
The theory of inflation postulates that the prompt expansion of Universe started just after Big Bang \cite{kubik_origin_2022, moore_big_2014}. After the inflationary era there must be numerous particles those are responsible for re-thermalization of Universe \cite{green_cosmological_2022}. The transition between inflation and further evolutionary phase of universe is reheating \cite{albrecht_inflation_1994,albrecht_reheating_1982,kofman_reheating_1994,allahverdi_reheating_2010}. This transition era is well parameterized using various reheating parameters \cite{cook_reheating_2015,yadav_reheating_2023,yadav2024reheating} and plays a crucial role in understanding standard matter production. The quantum properties and fluctuations related to inflaton provides an important information of early universe and gravity. Further, that it is related to classical gravity in Friedmann-Robertson-Walker (FRW) Universe \cite{mohajan_friedmann_2013, suresh_particle_2004}. According to cosmological principle Universe is homogeneous and isotropic. Cosmological, description of universe is based on Friedmann equations with scalar field(s). Under the classical gravity, Friedmann equations assume that those are valid even at initial stage of universe too. Though, quantum properties and quantum fluctuations of matter fields are anticipated to play a significant role. It is worth mentioning here that quantum effects into classical gravity can be better understood by Semi-Classical Theory of Gravity (SCTG). Further, the SCTG is based on fact that gravitational field depends upon quantized matter field having curved space-time. That has drawn special attention from theoretical physicists  \cite{kim_one-parameter_1999, finelli_quantum_1999, geralico_novel_2004, padmanabhan_gravity_2005} . Though quantum gravity effects can be consider negligible at this stage of universe. To describe the universe both gravity and matter fields are to be preserved quantum mechanically, but at present no consistent quantum theory is available to describe gravity. Hence, proper picture of early universe with a suitable cosmological model can be constructed in terms of the semi-classical Friedmann equations, where gravity can be preserved as classical with quantized matter field(s). The elementary idea of the Friedmann-Lema{\^ \i}tre-Robertson-Walker metric is based on particular solution of Einstein{'}s field equations of general relativity \cite{mahajan_particle_2008,lachieze-rey_cosmic_1995,ellis_cosmological_1998,carvalho_scalar_2004}. 

Concept of squeezed states was firstly introduced by Kennard as an analysis of Gaussian wave packet of harmonic oscillator \cite{bakke_geometric_2009}. The quantum effect of inflaton is based on semi-classical gravity in quantum optics considerations \cite{kennard_zur_1927,venkataratnam_particle_2004}. So the quantum mechanical number state
representation of scalar and complex scalar fields coupled to semi-classical FRW universe \cite{pedrosa_gaussian_2009, stoica_friedmann-lemaitre-robertson-walker_2016, hu_anisotropy_1978}, language of quantum optics is an important
to understand the quantum optical effects in cosmology \cite{dhayal_quantum_2020}. The number state representation was further analyzed
to identify the expanding universe and anisotropic universe \cite{fischetti_quantum_1979,hartle_quantum_1979,hartle_quantum_1980, hartle_quantum_1981}. Initial and final states of particle creation in expanding universe were also
described using number state \cite{anderson_effects_1983, campos_semiclassical_1994, geralico_novel_2004-1, pedrosa_exact_2007, lopes_gaussian_2009}.
Consider a suitable, quantum mechanical states of the universe such as Squeezed Vacuum State (SVS), Squeezed Number State (SNS), and Coherent Squeezed
Number States (CSNS) are used to understand the multifaceted nature of flat FRW Universe  \cite{kuo_semiclassical_1993, caves_quantum-mechanical_1981, matacz_coherent_1994, suresh_thermal_2001, suresh_squeezed_1998, suresh_nonclassical_2001, venkataratnam_nonclassical_2010,venkataratnam_density_2008, venkataratnam_oscillatory_2010, venkataratnam_behavior_2013, dhayal_quantum_2020-1, lachieze-rey_cosmological_1999,takahashi_thermo_1996, xu_quantum_2007}. \\ Quantum mechanical considerations of particle production due to Coherent Squeezed Number State (CSNS) and Squeezed Number States (SNS) in the oscillatory region of FRW universe under the light of semi-classical
gravity \cite{hartle_quantum_1980, hartle_quantum_1981}.
In this paper, we study density fluctuations for Squeezed Number State (SNS) and Coherent Squeezed Number States (CSNS) using quantum mechanical squeezed operator and displacement operator using number states.

 \section{Energy-Momentum Tensor in Semiclassical Gravity { }}

Various models of modern cosmology are constructed based upon classical gravity of Einstein field equations with scalar field on the FRW metric. Friedmann equations are further enumerated quantum mechanically for background metric. There is no significant description of quantum theory of gravity is available, so contemporary cosmological models, for background metric is considered as classical, while the matter field as quantum mechanical. Such type of theoretical models are called as semiclassical theory of gravity. In this circumstances, the description of the gravitational field using semi-classical theory of gravity, can be written as (here onwards we use the unit system c=$\hslash$=1 and {G}=$\frac{1}{\mathit{m}_{\mathit{p}}^2}$)

\begin{equation} \label{1}
\mathcal{E}_{\mu \nu }=\frac{8\pi }{\mathit{m}_{\mathit{p}}^2}\left\langle \mathcal{T}_{\mu \nu }\right\rangle.
\end{equation}

Where  \(\mathcal{E}_{\mu \nu }\)  is the Einstein tensor and \(\left\langle \mathcal{T}_{\mu \nu }\right\rangle\) is the expectation value of energy-momentum tensor in a suitable quantum state. The quantum state satisfy 
the time dependent Schrodinger equation and it can be written as  

\begin{equation} \label{2}
\mathit{i}\frac{ \partial }{\partial \mathit{t}}\psi  = \overset{\wedge }{\mathcal{H}}\psi,
\end{equation}
where \(\overset{\wedge }{\mathcal{H}}\) is Hamiltonian operator. Equation for Friedmann-Robertson-Walker (FRW) space-time having generalized coordinates \(\left(\mathit{r}_1,\mathit{r}_2,\right. \mathit{r}_3\),
\(\mathit{r}_4\)) can be written as

\begin{equation} \label{3}
\mathit{d}\mathit{s}^{2 }= -\mathit{d}\mathit{r}_4^2 + \mathcal{G}^{2 }(\mathbf{t})\left(\mathit{d}\mathit{r}_1^2 + \mathit{d}\mathit{r}_2^2 + \mathit{d}\mathit{r}_3^2\right),
\end{equation}
where $\mathcal{G}(\mathbf{t})$ is the scale factor.\\

Lagrangian density $\mathfrak{L}$ { }for massive inflaton for flat FRW universe is given as 

\begin{equation} \label{4}
\mathfrak{L} = -\frac{1}{2}\sqrt{(-\mathfrak{g})}\left(\mathfrak{g}^{\mu \nu }\partial _{\mu }\Phi  \partial _{\nu }\Phi + m^2\Phi  ^2\right).
\end{equation}

Klein-Gordon (K-G) equation for massive inflaton $\Phi$  can be derived from equations (\ref{3}) and (\ref{4}) as

\begin{equation} \label{5}
\ddot{\Phi  }+3\frac{\dot{\mathcal{G}} (\mathbf{t})}{\mathcal{G} (\mathbf{t})}\dot{\Phi  }+ m^2\Phi  =0,
\end{equation}
where Hubble parameter $\mathfrak{H}$=\(\frac{\dot{\mathcal{G}}(\mathbf{t})}{\mathcal{G}(\mathbf{t})}\) and \(\Phi \) is the Homogeneous scalar field for gravity to be consider, \(\dot{\Phi  }\) and \(\ddot{\Phi  }\) are the first and second order derivatives
with respect to time. $\Pi $ is the momentum conjugate to \(\overset{\wedge }\Phi \) { }is represented as 

\begin{equation} \label{6}
\Pi = \frac{ \partial \mathcal{L}}{\partial \dot\Phi}.
\end{equation}

Consider the canonical quantization, the Hamiltonian for the inflaton can be written as 

\begin{equation} \label{7}
{\mathcal{H}}=\frac{1}{2\mathcal{G}^3
(\mathbf{t})}\Pi ^2 +\frac{1}{2}\mathcal{G}^3 (\mathbf{t})m^{2 }\overset{\wedge }\Phi ^2,
\end{equation}

while temporal component of the energy-momentum tensor can be calculated as

\begin{equation} \label{8}
\mathcal{T}_{00}= \mathcal{G}^3 (\mathbf{t})\left(\frac{1}{2}\dot{\Phi }^2+\frac{1}{2}m^2\overset{\wedge }\Phi ^2\right) ,
\end{equation}

\section{Formulation of Squeezed Number State (SNS)}

Single mode squeezed coherent state is defined as
\begin{equation} \label{15}
|\varUpsilon ,\zeta \rangle =\overset{\wedge}W
(\rho ,\Psi )\mathfrak{D}(\varUpsilon )|0\rangle ,
\end{equation}
where $\mathfrak{D}(\varUpsilon)$ is displacement operator and is defined as 

\begin{equation} \label{26}
\mathfrak{D}(\Upsilon )=\exp (\Upsilon \overset{\wedge }{\mathit{e}} ^{\dagger }-\Upsilon ^*\overset{\wedge }{\mathit{e}}) ,
\end{equation}
and the squeezing operator \(\overset{\wedge}W(\rho ,\Psi )\) is given by
\begin{equation} \label{16}
\overset{\wedge}W(\rho ,\Psi )=\exp \frac{\rho }{2}\biggr(\exp (-\mathit{i}\Psi )\overset{\wedge }{\mathit{e}}^2-\exp (\mathit{i}\Psi)\overset{\wedge}{\mathit{e}}^ {\dagger 2 }\biggr) ,
\end{equation}
where $\rho $ can take values between 0 $\leq $ $\rho $ $\leq $ $\infty $ is squeezing parameter to regulates order of squeezing, $\Psi $ can take
values between -$\Pi $ $\leq $ $\Psi$ $\leq $ $\Pi $ { }is squeezing angle that governs distribution of conjugate variables. Squeezing operator have following property

\begin{equation} \label{17}
\overset{\wedge}W ^{\dagger }\overset{\wedge }{\mathit{e}}\overset{\wedge}W =\overset{\wedge }{\mathit{e}} \text{cosh$\rho $} - \overset{\wedge
}{\mathit{e}} ^{\dagger }\exp (\mathit{i}\Psi )\text{sinh$\rho $},
\end{equation}

\begin{equation} \label{18}
\overset{\wedge}W ^{\dagger }\overset{\wedge }{\mathit{e}}^{\dagger }\overset{\wedge}W =\overset{\wedge }{\mathit{e}}^{\dagger }\text{cosh$\rho
$} - \overset{\wedge }{\mathit{e}} \exp (-\mathit{i}\Psi )\text{sinh$\rho $}.
\end{equation}

Operation of squeezing parameter on number state gives as Squeezed Number State and it is defined as

\begin{equation} \label{21}
|\zeta ,n\rangle =\overset{\wedge}W(\rho ,\Psi )|n\rangle.
\end{equation}

The operation of annihilation operator \(\overset{\wedge }{\mathit{e}}\) and creation operator \(\overset{\wedge }{\mathit{e}} ^{\dagger }\) on number state is given as  

\begin{equation} \label{22}
\overset{\wedge }{\mathit{e}}|n\rangle =\sqrt{n}|n-1\rangle, \\
\end{equation}

\begin{equation} \label{23}
\overset{\wedge }{\mathit{e}} ^{\dagger }
|n\rangle =\sqrt{n+1}|n+1\rangle.
\end{equation}

Eigenstates for the Hamiltonian can be written as

\begin{equation} \label{24}
\overset{\wedge }{\mathit{e}} ^{\dagger }\overset{\wedge }{(\mathbf{t})\mathit{e}(\mathbf{t})}|n,\Phi ,\mathbf{t}\rangle =n|n,\Phi ,\mathbf{t}\rangle.
\end{equation}

\section{Formulation of Coherent Squeezed Number States (CSNS)}

Coherent Squeezed Number States (CSNS) is an important state in quantum optics. Single mode coherent state is descirbed as

\begin{equation} \label{25}
|\varUpsilon \rangle=\mathfrak{D}(\Upsilon )|0\rangle.
\end{equation}

The action of \(\overset{\wedge }{\mathit{e}}\) on the coherent state is given as 

\begin{equation} \label{27}
\overset{\wedge }{\mathit{e}}|\Upsilon \rangle =\Upsilon |\Upsilon \rangle.
\end{equation}

Annihilation operator \(\overset{\wedge }{\mathit{e}}\) and creation operator \(\overset{\wedge }{\mathit{e}} ^{\dagger }\) combined with single mode displacement operator $\mathfrak{D}$($\Upsilon $) which is given by Eq. (\ref{26}) and satisfy the following properties  

\begin{equation} \label{28}
\mathfrak{D}^{\dagger }\overset{\wedge }{\mathit{e}} ^{\dagger }\mathfrak{D}=\overset{\wedge }{\mathit{e}} ^{\dagger }+\Upsilon ^*,
\end{equation}

\begin{equation} \label{29}
\mathfrak{D}^{\dagger }\overset{\wedge }{\mathit{e}} \mathfrak{D}=\overset{\wedge }{\mathit{e}} +\Upsilon.
\end{equation}

Operation of single mode Displacement operator $\mathfrak{D}$($\Upsilon $) and squeezing parameter operator $\overset{\wedge}W(\rho ,\Psi )$ on Number State gives the coherent squeezed number state (CSNS) and it is defined as

\begin{equation} \label{31}
|\Upsilon ,\zeta ,n\rangle =\mathfrak{D}(\Upsilon )\overset{\wedge}W(\rho ,\Psi )|n\rangle .
\end{equation}

 \section{Oscillatory phase of inflaton and Semiclassical Einstein euquations for SNS and CSNS { }}

 Consider an oscillatory phase of massive inflaton minimally coupled to a flat FRW metric in a nonclassical states. Among  nonclassical states, firstly we consider it in squeezed Number State (SNS) and after that we consider it in coherent squeezed Number State (CSNS) in a flat FRW metric minimally coupled to massive inflaton. The temporal component of classical Einstein equation under classical gravity is written as

\begin{equation} \label{10}
\left(\frac{\dot{\mathcal{G}}(\mathbf{t})}{\mathcal{G}(\mathbf{t})}\right)^2=\frac{8\pi }{3\mathit{m}_{\mathit{p}}^2}\frac{\mathcal{T}_{00}}{\mathcal{G}^3
(\mathbf{t})},
\end{equation}

where, inflaton energy density \(\mathcal{T}_{00}\) is given by Eq. (\ref{8}). For the suitable quantum state with normal ordered expectation value of Hamiltonian \(\left\langle :\overset{\wedge }{\mathcal{H}_m}:\right\rangle\), Friedmann equation in the semiclassical theory of gravity can be written as

\begin{equation} \label{11}
\left(\frac{\dot{\mathcal{G}}(\mathbf{t})}{\mathcal{G}(\mathbf{t})}\right)^2=\frac{8\pi }{3\mathit{m}_{\mathit{p}}^2}\frac{1}{\mathcal{G}^3 (\mathbf{t})}\langle
:\overset{\wedge }{\mathcal{H}_m}:\rangle.
\end{equation}

We know that the annihilation operator (\(\overset{\wedge }{\mathit{e}}\)) and creation operator (\(\overset{\wedge }{\mathit{e}} ^{\dagger }\)) obey the commutation relation as 

\begin{equation} \label{32}
\left[\overset{\wedge }{\mathit{e}},\overset{\wedge }{\mathit{e}} ^{\dagger }\right]=1.
\end{equation}

Further, annihilation and creation operators can be computed as

\begin{equation} 
\overset{\wedge }{\mathit{e}}(\mathbf{t})=\Phi ^*(\mathbf{t})\overset{\wedge }\Pi -\mathcal{G}^3 (\mathbf{t})\dot{\Phi }^*(\mathbf{t})\overset{\wedge }\Phi, \\
\nonumber\\
\end{equation}

\begin{equation} \label{33}
\overset{\wedge }
{\mathit{e}} ^{\dagger }(\mathbf{t})=\Phi (\mathbf{t})\overset{\wedge }\Pi -\mathcal{G}^3 (\mathbf{t})\dot{\Phi }(\mathbf{t})\overset{\wedge }\Phi.
\end{equation}

Using Eqs.(\ref{24}, \ref{32}-\ref{33}), $\overset{\wedge }\Phi $ { }and $\Pi $ holds the relations as

\begin{equation} \label{34}
\overset{\wedge }\Phi =\frac{1}{i}\left(\Phi ^*\overset{\wedge }{\mathit{e}} ^{\dagger }-\Phi \overset{\wedge }{\mathit{e}} \right),\\
\end{equation}

\begin{equation} \label{35}
\overset{\wedge }\Phi ^2=\left(2\overset{\wedge }{\mathit{e}} ^{\dagger }\overset{\wedge }{\mathit{e}}+1\right)\Phi ^*\Phi -\left(\Phi ^*\overset{\wedge }{\mathit{e}}
^{\dagger }\right)^2-\left(\Phi \overset{\wedge }{\mathit{e}}\right)^2,
\end{equation}

\begin{equation} \label{36}
\overset{\wedge }\Pi =\text{i$\mathcal{G}$}^3 (\mathbf{t})\left(\dot{\Phi }\overset{\wedge }{\mathit{e}} -\dot{\Phi }^*\overset{\wedge }{\mathit{e}} ^{\dagger
} \right),\\
\end{equation}

\begin{equation} \label{37}
\overset{\wedge }\Pi ^2=\mathcal{G}^3 (\mathbf{t})\left[\left(2\overset{\wedge }{\mathit{e}} ^{\dagger }\overset{\wedge }{\mathit{e}}+1\right)\dot{\Phi }^*\dot{\Phi
}-\left(\dot{\Phi }^*\overset{\wedge }{\mathit{e}} ^{\dagger }\right)^2-\left(\dot{\Phi }\overset{\wedge }{\mathit{e}}\right)^2\right].
\end{equation}

Hamiltonian in semiclassical Friedmann equation computed using Eqs. (\ref{24}, \ref{11}-\ref{33}) for the number state as

\begin{equation} \label{12}
\left(\frac{\dot{\mathcal{G}}(\mathbf{t})}{\mathcal{G}(\mathbf{t})}\right)^2=\frac{8\pi }{3\mathit{m}_{\mathit{p}}^2}\left[\left(n+\frac{1}{2}\right)\left(\dot{\Phi
}^*(\mathbf{t})\dot{\Phi }(\mathbf{t})+m^2\Phi ^*(\mathbf{t})\Phi (\mathbf{t})\right)\right],
\end{equation}
where $\Phi $($\mathbf{t}$) and \(\Phi ^*(\mathbf{t})\) satisfy  the equation (\ref{5}) and the Wronskian condition given as 

\begin{equation} \label{13}
\mathcal{G}^3 (\mathbf{t})\biggr[\dot{\Phi }^*(\mathbf{t})\Phi (\mathbf{t})- \Phi ^*(\mathbf{t})\dot{\Phi }(\mathbf{t})\biggr]=\mathit{i}.
\end{equation}

Semiclassical Einstein equation for SNS and CSNS has an important consequences of above formulation.
To determine the Semiclassical Einstein equation for SNS, by using Eq. (\ref{16}-\ref{23}, \ref{37}) and after applying the SNS properties and computation  we get normal ordered expection value of $ \left\langle :\overset{\wedge }\Pi ^2:\right\rangle {} $ for Squeezed Number State is 

\begin{align} \label{101}
\left\langle :\overset{\wedge }\Pi ^2:\right\rangle {}_{\text{SNS}}=&2\mathcal{G}^6 (\mathbf{t})\biggr[\biggr(n\text{Cosh}^2\rho +(n+\frac{1}{2})\text{Sinh}^2\rho +\frac{1}{2}\biggr)\dot{\Phi }^*\dot{\Phi }\nonumber\\
&-\biggr((n+\frac{1}{2})\text{Sinh$\rho $\text{Cosh$\rho $}}\biggr)\dot{\Phi }^{*2}-\biggr((n+\frac{1}{2})\text{Sinh$\rho $\text{Cosh$\rho $}}\biggr)\dot{\Phi }^2\biggr].
\end{align}

using Eqs. (\ref{16}-\ref{23}, \ref{35}) and after computation, we get the normal ordered expectation value of $ \left\langle :\overset{\wedge }\Phi ^2:\right\rangle {} $ for Squeezed Number State (SNS) is

\begin{align} \label{102}
\left\langle :\overset{\wedge }\Phi ^2:\right\rangle {}_{\text{SNS}}=&2\biggr[\biggr(n\text{Cosh}^2\rho +(n+\frac{1}{2})\text{Sinh}^2\rho +\frac{1}{2}\biggr){\Phi }^*{\Phi }\nonumber\\
&-\biggr((n+\frac{1}{2})\text{Sinh$\rho $\text{Cosh$\rho $}}\biggr){\Phi }^{*2}-\biggr((n+\frac{1}{2})\text{Sinh$\rho $\text{Cosh$\rho $}}\biggr){\Phi }^2\biggr].
\end{align}

using Eqs.(\ref{7}, \ref{101}-\ref{102}), we compute the normal ordered expectation value of the Hamiltonian in Squeezed Number State (SNS) as  

\begin{align} \label{103}
\left\langle :\overset{\wedge }{\mathcal{H}}:\right\rangle_{\text{SNS}}=&\mathcal{G}^3 (\mathbf{t})\biggr[\biggr(n\text{Cosh}^2\rho +(n+\frac{1}{2})\text{Sinh}^2\rho +\frac{1}{2}\biggr)\biggr(\dot{\Phi }^*\dot{\Phi }+m^2{\Phi }^*{\Phi }\biggr)\nonumber\\
& -\biggr((n+\frac{1}{2})\text{Sinh$\rho $\text{Cosh$\rho $}}\biggr)\biggr(\dot{\Phi }^{*2}+m^2{\Phi }^{*2}\biggr)\nonumber\\
&-\biggr((n+\frac{1}{2})\text{Sinh$\rho $\text{Cosh$\rho $}}\biggr)\biggr(\dot{\Phi^2}+m^2{\Phi^2}\biggr)\biggr].
\end{align}

using Eqs.(\ref{103}) in Eqs.(\ref{11}), the semiclassical Einstein equation for Squeezed Number State (SNS) is 

\begin{align} \label{104}
\left(\frac{\dot{\mathcal{G}}(\mathbf{t})}{\mathcal{G}(\mathbf{t})}\right)^2_{\text{SNS}}=&\frac{8\pi }{3\mathit{m}_{\mathit{p}}^2}\biggr[\biggr(n\text{Cosh}^2\rho +(n+\frac{1}{2})\text{Sinh}^2\rho +\frac{1}{2}\biggr)\biggr(\dot{\Phi }^*\dot{\Phi }+m^2{\Phi }^*{\Phi }\biggr)\nonumber\\
& -\biggr((n+\frac{1}{2})\text{Sinh$\rho $\text{Cosh$\rho $}}\biggr)\biggr(\dot{\Phi }^{*2}+m^2{\Phi }^{*2}\biggr)\nonumber\\
&-\biggr((n+\frac{1}{2})\text{Sinh$\rho $\text{Cosh$\rho $}}\biggr)\biggr(\dot{\Phi^2}+m^2{\Phi^2}\biggr)\biggr].
\end{align}

Now, we determine the Semiclassical Einstein Equation for Coherent Squeezed Nunber State (CSNS), by using Eq. (\ref{16}-\ref{31}, \ref{37}) and after applying the CSNS properties and computation we get normal ordered expection value of $ \left\langle :\overset{\wedge }\Pi ^2:\right\rangle {} $ for Coherent Squeezed Number State (CSNS) is 
\begin{align} \label{105}
\langle :\overset{\wedge }\Pi^2 :\rangle {}_{\text{CSNS}}=&2\mathcal{G}^6 (\mathbf{t})\biggl[\biggl(n\cosh^2\rho +(n+\frac{1}{2})\sinh^2\rho +\frac{1}{2}+\varUpsilon^* \varUpsilon \biggr)\dot{\Phi}^*\dot{\Phi}\nonumber\\
&-\biggl((n+\frac{1}{2})\sinh\rho\cosh\rho -\frac{\varUpsilon^{*2}}{2}\biggr)\dot{\Phi}^{*2}\nonumber\\
&-\biggl((n+\frac{1}{2})\sinh\rho\cosh\rho -\frac{\varUpsilon^2}{2}\biggr)\dot{\Phi}^2\biggr].
\end{align}

using Eqs. (\ref{16}-\ref{31}, \ref{35}) after computation, we get normal ordered expection value of $ \left\langle :\overset{\wedge }\Phi ^2:\right\rangle {} $ for Coherent Squeezed Number State (CSNS) is

\begin{align} \label{106}
\left\langle :\overset{\wedge }\Phi ^2:\right\rangle {}_{\text{CSNS}}=&2\biggr[\biggr(n\text{Cosh}^2\rho +(n+\frac{1}{2})\text{Sinh}^2\rho +\frac{1}{2}+\varUpsilon ^*\varUpsilon\biggr){\Phi }^*{\Phi }\nonumber\\
&-\biggr((n+\frac{1}{2})\text{Sinh$\rho $\text{Cosh$\rho $}}-\frac{\varUpsilon ^*2}{2}\biggr){\Phi }^{*2}\nonumber\\
&-\biggr((n+\frac{1}{2})\text{Sinh$\rho $\text{Cosh$\rho $}}-\frac{\varUpsilon ^2}{2}\biggr){\Phi }^2\biggr].
\end{align}

using Eqs.(\ref{7}, \ref{105}-\ref{106}), we get the normal ordered expectation value of Hamiltanian in Coherent Squeezed Number State (CSNS) as 

\begin{align} \label{107}
\left\langle :\overset{\wedge }{\mathcal{H}}:\right\rangle_{\text{CSNS}}=&\mathcal{G}^3 (\mathbf{t})\biggr[\biggr(n\text{Cosh}^2\rho +(n+\frac{1}{2})\text{Sinh}^2\rho +\frac{1}{2}+\varUpsilon ^*\varUpsilon\biggr)\biggr(\dot{\Phi }^*\dot{\Phi }+m^2{\Phi }^*{\Phi }\biggr)\nonumber\\
& -\biggr((n+\frac{1}{2})\text{Sinh$\rho $\text{Cosh$\rho $}}-\frac{\varUpsilon ^{*2}}{2}\biggr)\biggr(\dot{\Phi }^{*2}+m^2{\Phi }^{*2}\biggr)\nonumber\\
&-\biggr((n+\frac{1}{2})\text{Sinh$\rho $\text{Cosh$\rho $}}-\frac{\varUpsilon ^2}{2}\biggr)\biggr(\dot{\Phi^2}+m^2{\Phi^2}\biggr)\biggr].
\end{align}

using Eqs.(\ref{107}) in Eqs.(\ref{11}), the semiclassical Einstein equation for Coherent Squeezed Number State (CSNS) is 

\begin{align} \label{1041}
\left(\frac{\dot{\mathcal{G}}(\mathbf{t})}{\mathcal{G}(\mathbf{t})}\right)^2_{\text{CSNS}}=&\frac{8\pi }{3\mathit{m}_{\mathit{p}}^2}\biggr[\biggr(n\text{Cosh}^2\rho +(n+\frac{1}{2})\text{Sinh}^2\rho +\frac{1}{2}+\varUpsilon ^*\varUpsilon\biggr)\biggr(\dot{\Phi }^*\dot{\Phi }+m^2{\Phi }^*{\Phi }\biggr)\nonumber\\
& -\biggr((n+\frac{1}{2})\text{Sinh$\rho $\text{Cosh$\rho $}}-\frac{\varUpsilon ^{*2}}{2}\biggr)\biggr(\dot{\Phi }^{*2}+m^2{\Phi }^{*2}\biggr)\nonumber\\
&-\biggr((n+\frac{1}{2})\text{Sinh$\rho $\text{Cosh$\rho $}}-\frac{\varUpsilon ^2}{2}\biggr)\biggr(\dot{\Phi^2}+m^2{\Phi^2}\biggr)\biggr].
\end{align}

\section{Density fluctuations for Oscillatory phase of inflaton for Squeezed Number State and Coherent Squeezed Number State}

The semiclassical theory of FRW universe predicts fluctuations for energy momentum tensor. The energy momentum tensor of the inflaton for quantum states related to particle production in FRW universe can be studied with the help of semiclassical Einstein equation. Semiclassical Einstein equation for flat FRW universe can be determine using following relation

\begin{equation} \label{38}
\triangle = \langle :T_{\mu \nu }^2:\rangle  - \langle :T_{\mu \nu }:\rangle ^2.
\end{equation}

Here \(\left\langle :T_{\mu \nu }^2:\rangle \right.\) and \(\left\langle :T_{\mu \nu }:\rangle ^2 \right.\) represents normal ordered expectation value of square of energy
momentum tensor and square of the normal ordered expectation value of energy momentum tensor respectively. Symbol : : used to demonstrate that normal ordered value of
physical quantities are to be computed with respect to normal ordering of scalar field. In this study we have determined density fluctuations for Squeezed
Number State and Coherent Squeezed Number State using single mode temporal component of the energy momentum tensor. 

\subsection{Density fluctuations for Squeezed Number State (SNS)}

Density fluctuations for Squeezed Number State is calculated using Eq. (\ref{38}),normal ordered expectation value of square of energy momentum tensor \(\left\langle :T_{00}^2:\right\rangle  \text {}\) for Squeezed Number State is given as

\begin{align} \label{39}
\left\langle :T_{00}^2:\right\rangle_{\text{SNS}} =&\frac{1}{4}\mathcal{G}^6 (\mathbf{t})m^{4 }\left\langle :\overset{\wedge }\Phi ^4:\right\rangle_{\text{SNS}}+\frac{m^{2
}}{4}\left\langle :\overset{\wedge }\Phi ^2\overset{\wedge }\Pi ^2:\right\rangle {}_{\text{SNS}}\nonumber\\
&+\frac{m^{2 }}{4}\left\langle :\overset{\wedge }\Pi ^2\overset{\wedge }\Phi ^2:\right\rangle_{\text{SNS}}+\frac{1}{4\mathcal{G}^6
(\mathbf{t})}\left\langle :\overset{\wedge }\Pi ^4:\right\rangle_{\text{SNS}}.
\end{align}

Using Eqs. (\ref{17}-\ref{18}, \ref{32}-\ref{37}) the values of \(\left\langle :\overset{\wedge }\Pi ^4:\right\rangle {}_{\text{SNS}}\) can be computed as

\begin{align} \label{41}
\left\langle :\overset{\wedge }\Pi ^4:\right\rangle _{\text{SNS}}=&\frac{\mathcal{G}^6(\mathbf{t})}{4m^{2 }\mathbf{t}^4}\biggr[3+\text{Cosh}^3\rho\text{Sinh$\rho
$}\left(24n^2\right)\nonumber\\
&+\text{Cosh$\rho $} \text{Sinh}^3\rho \left(24n^2+48n+24\right)\nonumber\\
&+\text{Cosh}^2\rho\text{Sinh}^2\rho \left(36n^2+36n+12\right)\nonumber\\
&+\text{Cosh$\rho
$} \text{Sinh$\rho $}(24n+12)+\text{Sinh}^4\rho \left(6n^2+18n+12\right)\nonumber\\
&+\text{Sinh}^2\rho (12n+12)+\text{Cosh}^4\rho (6n^2-6n)+\text{Cosh}^2\rho
(12n)\biggr].
\end{align}

Using Eqs. (\ref{17}-\ref{18}, \ref{32}-\ref{37}) the values of \(\left\langle :\overset{\wedge }\Phi ^4:\right\rangle _{\text{SNS}}\) can be computed as

\begin{align} \label{43}
\left\langle :\overset{\wedge }\Phi ^4:\right\rangle _{\text{SNS}}=&\frac{1}{4m^{2 }\mathcal{G}^6(\mathbf{t})}
\biggr[3+\text{Cosh}^3\rho\text{Sinh$\rho $}\left(24n^2\right)+\text{Cosh$\rho
$} \text{Sinh}^3\rho \left(24n^2\right.\nonumber\\
&+48n+24)+\text{Cosh}^2\rho\text{Sinh}^2\rho \left(36n^2+36n+12\right)\nonumber\\
&+\text{Cosh$\rho $} \text{Sinh$\rho
$}(24n+12)+\text{Sinh}^4\rho \left(6n^2+18n+12\right)\nonumber\\
&+\text{Sinh}^2\rho (12n+12)+\text{Cosh}^4\rho (6n^2-6n)+\text{Cosh}^2\rho (12n)\biggr].
\end{align}

Using Eqs. (\ref{17}-\ref{18}, \ref{32}-\ref{37}) the values of \(\left\langle :\overset{\wedge }\Phi ^2\overset{\wedge }\Pi ^2:\right\rangle _{\text{SNS}}\) can be computed as 

\begin{align} \label{45}
\left\langle :\overset{\wedge }\Phi ^2\overset{\wedge }\Pi ^2:\right\rangle _{\text{SNS}}=&\frac{1}{4m^{2 }\mathbf{t}^2}\biggr[3+\text{Cosh}^3\rho \text{Sinh$\rho $}\left(24n^2\right)+\text{Cosh$\rho
$} \text{Sinh}^3\rho \left(24n^2\right)\nonumber\\
&+48n+24)+\text{Cosh}^2\rho \text{Sinh}^2\rho \left(36n^2+36n+12\right)\nonumber\\
&+\text{Cosh$\rho $} \text{Sinh$\rho
$}(24n+12)+\text{Sinh}^4\rho \left(6n^2+18n+12\right)\nonumber\\
&+\text{Sinh}^2\rho (12n+12)+\text{Cosh}^4\rho (6n^2-6n)\nonumber\\
&+\text{Cosh}^2\rho (12n)\biggr].
\end{align}

Using Eqs. (\ref{17}-\ref{18}, \ref{32}-\ref{37}) the values of \(\left\langle :\overset{\wedge }\Pi ^2\overset{\wedge }\Phi ^2:\right\rangle {}_{\text{SNS}}\) can be computed as 

\begin{align} \label{47}
\left\langle:\overset{\wedge }\Pi^2\overset{\wedge }\Phi^2:\right\rangle_{\text{SNS}}=&\frac{1}{4m^{2}\mathbf{t}^2}\biggr[3+\text{Cosh}^3\rho\text{Sinh$\rho $}\left(24n^2\right)+
\text{Cosh$\rho$}\text{Sinh}^3\rho\left(24n^2\right)
\nonumber\\
&+48n+24)+\text{Cosh}^2\rho\text{Sinh}^2\rho \left(36n^2+36n+12\right)\nonumber\\
&+\text{Cosh$\rho$}\text{Sinh$\rho
$}(24n+12)+\text{Sinh}^4\rho\left(6n^2+18n+12\right)
\nonumber\\
&+\text{Sinh}^2\rho(12n+12)+\text{Cosh}^4\rho(6n^2-6n)\nonumber\\&+\text{Cosh}^2\rho(12n)\biggr].
\end{align}

Substituting { }the values of \(\left\langle :\overset{\wedge }\Phi ^4:\right\rangle {}_{\text{SNS}}\), \(\left\langle :\overset{\wedge }\Pi ^4:\right\rangle {}_{\text{SNS}}\), \(\left\langle
:\overset{\wedge }\Phi ^2\overset{\wedge }\Pi ^2:\right\rangle {}_{\text{SNS}}\), \(\left\langle :\overset{\wedge }\Pi ^2\overset{\wedge }\Phi ^2:\right\rangle {}_{\text{SNS}}\) in Eq. (\ref{39}), the normal ordered expection value of square of energy momentum tensor can be calculated as

\begin{align} \label{48}
\left\langle :T_{00}^2:\right\rangle_{\text{SNS}}=&\left(\frac{1}{16m^{2 }\mathbf{t}^4}+\frac{1}{8\mathbf{t}^2}+\frac{m^{2}}{16}\right)
\biggr[3+\text{Cosh}^3\rho
 \text{Sinh$\rho$}(24n^2\nonumber\\
&+\text{Cosh$\rho$}\text{Sinh}^3\rho \left(24n^2+48n+24\right)+\text{Cosh}^2\rho  \text{Sinh}^2\rho(36n^2
\nonumber\\
&\left.+36n+12\right)+\text{Cosh$\rho
$}\text{Sinh$\rho $}(24n+12)+\text{Sinh}^4\rho (6n^2
\nonumber\\
&\left.+18n+12\right)+\text{Sinh}^2\rho (12n+12)+\text{Cosh}^4\rho (6n^2-6n)\nonumber\\
&+\text{Cosh}^2\rho
(12n)\biggr].
\end{align}

Now, normal ordered expectation value of temporal component of energy momentum tensor is

\begin{equation} \label{49}
\left\langle :T_{00}:\right\rangle _{\text{SNS}} =\frac{1}{2}\mathcal{G}^3 (\mathbf{t})m^{2 }\left\langle :\overset{\wedge }\Phi ^2:\right\rangle {}_{\text{SNS}}+\frac{1}{2\mathcal{G}^3
(\mathbf{t})}\left\langle :\overset{\wedge }\Pi ^2:\right\rangle {}_{\text{SNS}}.
\end{equation}

Using Eqs. (\ref{17}-\ref{18}, \ref{32}-\ref{37}) the values of \(\left\langle :\overset{\wedge }\Pi ^2:\right\rangle {}_{\text{SNS}}\) can be computed as

\begin{equation} \label{51}
\left\langle :\overset{\wedge }\Pi ^2:\right\rangle {}_{\text{SNS}}=\frac{\mathcal{G}^3 (\mathbf{t})}{2m\mathbf{t}^2}\biggr[\text{Cosh}^2\rho (2n)+\text{Sinh}^2\rho (2n+2)+1
+\text{Cosh$\rho $} \text{Sinh$\rho $}(4n+2)\biggr].
\end{equation}

Using Eqs. (\ref{17}-\ref{18}, \ref{32}-\ref{37}) the values of \(\left\langle :\overset{\wedge }\Phi ^2:\right\rangle {}_{\text{SNS}}\) can be computed as

\begin{align} \label{53}
\left\langle :\overset{\wedge }\Phi ^2:\right\rangle_{\text{SNS}}=
&\frac{1}{2\text{m$\mathcal{G}$}^3 (\mathbf{t})}\biggr[\text{Cosh}^2\rho (2n)+\text{Sinh}^2\rho (2n+2)+1
\nonumber\\
&+\text{Cosh$\rho $} \text{Sinh$\rho $}(4n+2)\biggr],
\end{align}
using Eqs. (\ref{51}) and (\ref{53}) in Eq.(\ref{49}), normal ordered value of temporal component of energy momentum tensor will be 

\begin{align} \label{54}
\left\langle :T_{00}:\right\rangle _{\text{SNS}}=&\left[\frac{m}{4}+\frac{1}{4\text{m$\mathbf{t}$}^2}\right]\biggr[\text{Cosh}^2\rho (2n)+\text{Sinh}^2\rho
(2n+2)+1\nonumber\\
&+\text{Cosh$\rho $} \text{Sinh$\rho $}(4n+2)\biggr].
\end{align}

Square of normal ordered value of temporal component of energy momentum tensor can be computed as

\begin{align} \label{55}
\left \langle :T_{00}: \right\rangle^2_{\text{SNS}}= & 
\left(\frac{1}{16m^{2 }\mathbf{t}^4}+\frac{1}{8\mathbf{t}^2}+\frac{m^{2 }}{16}\right)
\biggr[
1+\text{Cosh}^3\rho\text{Sinh$\rho $}\left(16n^2+8n\right)\nonumber \\
& +\text{Cosh$\rho $} \text{Sinh}^3\rho \left(16n^2+24n+8\right)+\text{Cosh}^2\rho  \text{Sinh}^2\rho \left(24n^2 \right. \nonumber\\
& \left.+24n+4\right)+\text{Cosh$\rho
$} \text{Sinh$\rho $}(8n+4)+\text{Sinh}^4\rho \left(4n^2\right. \nonumber\\
&+8n+4)+\text{Sinh}^2\rho (4n+4)+\text{Cosh}^4\rho \left(4n^2\right)\nonumber\\
&+\text{Cosh}^2\rho
(4n)\biggr].
\end{align}

Substituting the values of \(\left\langle :T_{00}^2:\right\rangle {}_{\text{SNS}}\) and \(\left\langle :T_{00}:\rangle \right.{}^2{}_{\text{SNS}}\)
in Eq. (\ref{38}), density fluctuations for Squeezed Number State (SNS) is 

\begin{align} \label{56}
\triangle_{\text{SNS}}=
&\left(\frac{1}{16m^{2}\mathbf{t}^4}+\frac{1}{8\mathbf{t}^2}+\frac{m^{2 }}{16}\right)
\biggr[2+\text{Cosh}^3\rho \text{Sinh$\rho
$}\left(8n^2-8n\right)\nonumber\\
&\text{Cosh$\rho $}\text{Sinh}^3\rho \left(8n^2+24n+16\right)+\text{Cosh}^2\rho  \text{Sinh}^2\rho\left(12n^2\right.\nonumber\\
&\left. 12n+8\right)+\text{Cosh$\rho$}\text{Sinh$\rho$}(16n+8)+\text{Sinh}^4\rho \left(2n^2+10n+8\right)\nonumber\\
&+\text{Sinh}^2\rho (8n+8)+\text{Cosh}^4\rho (2n^2-6n)+\text{Cosh}^2\rho (8n)\biggr].
\end{align}

\begin{figure}[h]%
\centering
\includegraphics[width=0.9\textwidth]{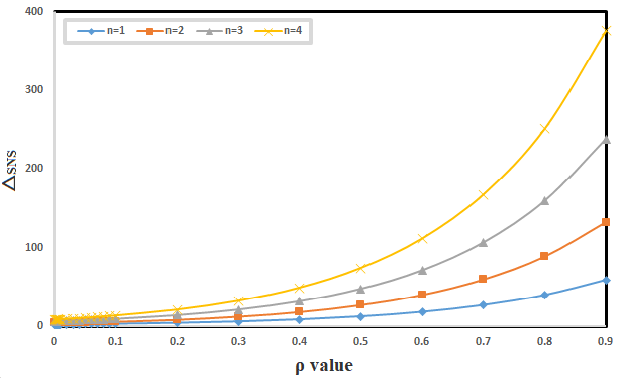}
\caption{Variation of density fluctuations \(\triangle _{\text{SNS}}\) for various values of squeezing parameter $\rho $}
\label{fig:figure_1}
\end{figure}

\begin{figure}[h]%
\centering
\includegraphics[width=0.9\textwidth]{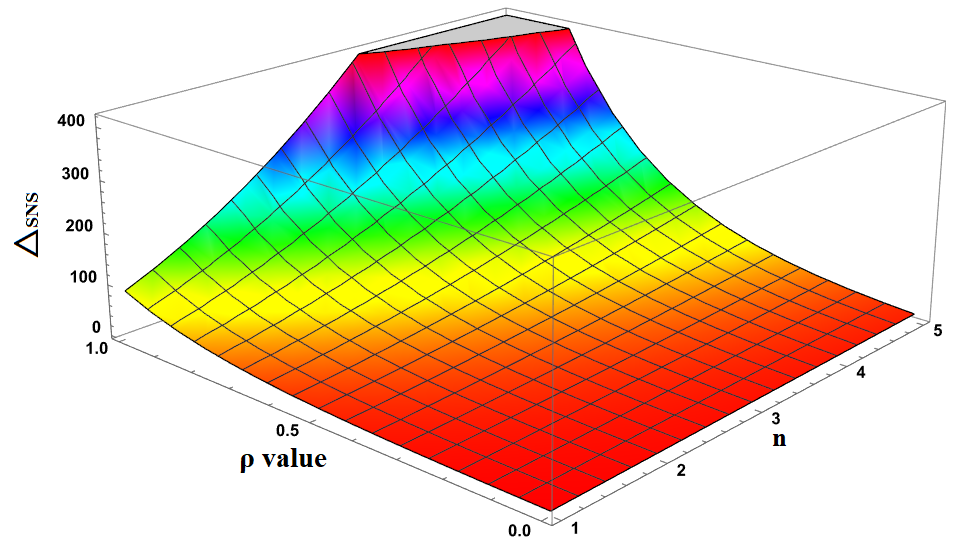}
\caption{3-D plot showing variation of density fluctuations \(\triangle _{\text{SNS}}\) with the values of n and squeezing parameter
$\rho $}
\label{fig:figure_2}
\end{figure}

Eq. (\ref{56}) shows the \(\triangle _{\text{SNS}}\) density fluctuations for squeezed number state depend on squeezing parameter ($\rho $) as well
as number state (n), but it is more strongly depend on number state rather than the squeezing parameter ($\rho $). Numerical
values of density fluctuations \(\triangle _{\text{SNS}}\) for various values of squeezed number state can be calculated using Eq. (\ref{56}), for n=1, 2, 3, 4 is as shown in table 1, 2 , 3, while assuming m=$\mathbf{t}$=1. Table \ref{tab:table_1} and \ref{tab:table_2} shows, for $\rho $ ranging between 0 .001 to 0.09 the variation in density fluctuations
is very small for all number state taken into consideration, but table \ref{tab:table_3} shows that, for $\rho $ ranging 0.1 to 0.09 shows large variation in \(\triangle
_{\text{SNS}}\). As we take n=0 in Eq. (\ref{56}), it reduces into density fluctuations for squeezed vacuum state \cite{ venkataratnam_density_2008, venkataratnam_oscillatory_2010, venkataratnam_behavior_2013}, showing the validity of approximation
used in SCEE. Fig. \ref{fig:figure_1}  shows variation of density fluctuations \(\triangle _{\text{SNS}}\), for various values of squeezed number state with squeezing parameter
$\rho $. It can be deduced from Fig. \ref{fig:figure_1} that with increasing value of $\rho $ as well as n, density fluctuations \(\triangle _{\text{SNS}}\) increases.
 In Fig.  \ref{fig:figure_2}, 3-D plot between $\rho $, n and \(\triangle _{\text{SNS}}\) shows variation of density fluctuations \(\triangle _{\text{SNS}}\) with various number
state and squeezing parameter $\rho $
\begin{table}[h]\label{table 1}
\caption{Numerical values of density fluctuations \(\triangle _{\text{SNS}}\) for various squeezed number state for squeezing parameter $\rho <<<$1}\label{t1}%
\label{tab:table_1}
\begin{tabular}{@{}llllll@{}}
\toprule
$\rho$  & n=1  & n=2 & n=3 & n=4\\
\midrule
0.001 & 2.50601 & 4.51403 & 6.52605 & 8.54208 \\
 0.002 & 2.51206 & 4.52812 & 6.55221 & 8.58432 \\
 0.003 & 2.51813 & 4.54227 & 6.57847 & 8.62672 \\
 0.004 & 2.52423 & 4.55648 & 6.60484 & 8.66929 \\
 0.005 & 2.53035 & 4.57075 & 6.63131 & 8.71201 \\
 0.006 & 2.53651 & 4.58509 & 6.65789 & 8.75490 \\
 0.007 & 2.54269 & 4.59948 & 6.68457 & 8.79796 \\
 0.008 & 2.54890 & 4.61394 & 6.71136 & 8.84118 \\
 0.009 & 2.55515 & 4.62846 & 6.73826 & 8.88456 \\
\botrule
\end{tabular}  
\end{table}

\begin{table}[h]\label{table 2}
\caption{ Numerical values of density fluctuations \(\triangle _{\text{SNS}}\) for various squeezed number state for squeezing parameter $\rho <<$1}\label{t2}%
\label{tab:table_2}
\begin{tabular}{@{}llllll@{}}
\toprule
$\rho$  & n=1  & n=2 & n=3 & n=4\\
\midrule
 0.01 & 2.56142 & 4.64304 & 6.76527 & 8.92811 \\
 0.02 & 2.62573 & 4.79231 & 7.04137 & 9.37291 \\
 0.03 & 2.69305 & 4.94804 & 7.32873 & 9.83511 \\
 0.04 & 2.76347 & 5.11049 & 7.62782 & 10.31550 \\
 0.05 & 2.83711 & 5.27992 & 7.93912 & 10.81470 \\
 0.06 & 2.91410 & 5.45659 & 8.26312 & 11.33370 \\
 0.07 & 2.99453 & 5.64079 & 8.60034 & 11.87320 \\
 0.08 & 3.07856 & 5.83282 & 8.95133 & 12.43410 \\
 0.09 & 3.16630 & 6.03296 & 9.31664 & 13.01730 \\
\botrule
\end{tabular}
\label{tab:msg1}  
\end{table}
\begin{table}[h]\label{table 3}
\caption{ Numerical values of density fluctuations \(\triangle _{\text{SNS}}\) for various squeezed number state for squeezing parameter $\rho <$1}\label{t3}%
\label{tab:table_3}
\begin{tabular}{@{}llllll@{}}
\toprule
$\rho$  & n=1  & n=2 & n=3 & n=4\\
\midrule
 0.1 & 3.25790 & 6.24150 & 9.69680 & 13.62380 \\
 0.2 & 4.42100 & 8.87210 & 14.46600 & 21.20270 \\
 0.3 & 6.17420 & 12.81450 & 21.58080 & 32.47310 \\
 0.4 & 8.79540 & 18.70150 & 32.19470 & 49.27500 \\
 0.5 & 12.70040 & 27.47850 & 48.02890 & 74.35150 \\
 0.6 & 18.50970 & 40.55610 & 71.65060 & 111.79300 \\
 0.7 & 27.14900 & 60.03830 & 106.89000 & 167.70500 \\
 0.8 & 39.99840 & 89.06340 & 159.46100 & 251.19200 \\
 0.9 & 59.11520 & 132.31200 & 237.88900 & 375.84600\\
\botrule
\end{tabular}
\end{table}

\subsection{Density fluctuations for Coherent Squeezed Number States (CSNS)}

Density fluctuations for coherent squeezed number state is being calculated using Eq. (\ref{38}), where normal ordered expectation value of square of energy momentum tensor \(\left\langle :T_{00}^2:\right\rangle {}_{\text{CSNS}}\) for squeezed number state is given as

\begin{align} \label{57}
\left\langle :T_{00}^2:\right\rangle_{\text{CSNS}}=
&\frac{1}{4}\mathcal{G}^6(\mathbf{t})m^{4}\left\langle :\overset{\wedge }\Phi ^4:\right\rangle_{\text{CSNS}}\nonumber\\
&+\frac{m^{2
}}{4}\left\langle:\overset{\wedge }\Phi ^2\overset{\wedge }\Pi ^2:\right\rangle _{\text{CSNS}}+\frac{m^{2 }}{4}\left\langle :\overset{\wedge }\Pi ^2\overset{\wedge }\Phi ^2:\right\rangle_{\text{CSNS}}\nonumber\\
&+\frac{1}{4\mathcal{G}^6
(\mathbf{t})}\left\langle :\overset{\wedge }\Pi ^4:\right\rangle_{\text{CSNS}}.
\end{align}

Using Eqs. (\ref{17}-\ref{18}, \ref{28}-\ref{37}) the values of \(\left\langle :\overset{\wedge }\Pi ^4:\right\rangle {}_{\text{CSNS}}\) can be computed as

\begin{align} \label{60}
\left\langle :\overset{\wedge }\Pi ^4:\right\rangle_{\text{CSNS}}=
&\frac{\mathcal{G}^6(\mathbf{t})}{4m^{2 }\mathbf{t}^4}\biggr[3+\Upsilon ^{*4}+\Upsilon^4+6\Upsilon
^{*2}\Upsilon ^2+12\Upsilon ^*\Upsilon -6\Upsilon ^{*2}-6\Upsilon ^2\nonumber\\
&-4\Upsilon ^{*3}\Upsilon -4\Upsilon ^*\Upsilon ^3+\text{Cosh}^4\rho (6n^2-6n)\nonumber\\
&+\text{Sinh}^4\rho
\left(6n^2+18n+12\right)+\text{Cosh}^2\rho\text{Sinh}^2\rho \left(36n^2+36n+12\right)\nonumber\\
&+\text{Cosh}^3\rho \text{Sinh$\rho $}\left(24n^2\right)+\text{Cosh$\rho
$} \text{Sinh}^3\rho \left(24n^2+48n+24\right)\nonumber\\
&+\text{Cosh$\rho $} \text{Sinh$\rho $}(24n+12)\left\{1+2\Upsilon ^*\Upsilon -\Upsilon ^{*2}-\Upsilon
^2\right\}\nonumber\\
&+\text{Sinh}^2\rho (12n+12)\left\{1+2\Upsilon ^*\Upsilon -\Upsilon^{*2}-\Upsilon ^2\right\}\nonumber\\
&+\text{Cosh}^2\rho (12n)\left\{1+2\Upsilon ^*\Upsilon
-\Upsilon ^{*2}-\Upsilon ^2\right\}\biggr].
\end{align}

Using Eqs. (\ref{17}-\ref{18}, \ref{28}-\ref{37}) the values of \(\left\langle :\overset{\wedge }\Phi ^4:\right\rangle _{\text{CSNS}}\) can be computed as

\begin{align} \label{63}
\left\langle :\overset{\wedge }\Phi ^4:\right\rangle_{\text{CSNS}} = &\frac{1}{4m^{2 }\mathcal{G}^6(\mathbf{t})} \biggr[3+\Upsilon ^{*4}+\Upsilon ^4+6\Upsilon ^{*2}\Upsilon
^2+12\Upsilon ^*\Upsilon\nonumber\\
&-6\Upsilon ^{*2}-6\Upsilon ^2-4\Upsilon ^{*3}\Upsilon -4\Upsilon ^*\Upsilon ^3+\text{Cosh}^4\rho \left(6n^2-6n\right)\nonumber\\
&+\text{Sinh}^4\rho
\left(6n^2+18n+12\right)+\text{Cosh}^2\rho\text{Sinh}^2\rho \left(36n^2+36n+12\right)\nonumber\\
&+\text{Cosh}^3\rho\text{Sinh$\rho $}\left(24n^2\right)+\text{Cosh$\rho
$} \text{Sinh}^3\rho \left(24n^2+48n+24\right)\nonumber\\
&+\text{Cosh$\rho $} \text{Sinh$\rho $}(24n+12)\left\{1+2\Upsilon ^*\Upsilon\right\}\nonumber\\
&-\Upsilon ^{*2}-\Upsilon
^2+\text{Sinh}^2\rho (12n+12)\left\{1+2\Upsilon ^*\Upsilon -\Upsilon ^{*2}-\Upsilon ^2\right\}\nonumber\\
&+\text{Cosh}^2\rho(12n)\left\{1+2\Upsilon ^*\Upsilon
-\Upsilon ^{*2}-\Upsilon ^2\right\}\biggr].
\end{align}

Using Eqs. (\ref{17}-\ref{18}, \ref{28}-\ref{37}) the values of \(\left\langle :\overset{\wedge }\Phi ^2\overset{\wedge }\Pi ^2:\right\rangle_{\text{CSNS}}\) can be calculated as

\begin{align} \label{66}
\left\langle :\overset{\wedge }\Phi ^2\overset{\wedge }\Pi ^2:\right\rangle _{\text{CSNS}}=&\frac{1}{4m^{2 }\mathbf{t}^2}\biggr[3+\Upsilon ^{*4}+\Upsilon ^4+6\Upsilon ^{*2}\Upsilon
^2+12\Upsilon ^*\Upsilon -6\Upsilon ^{*2}\nonumber\\
&-6\Upsilon ^2-4\Upsilon ^{*3}\Upsilon -4\Upsilon ^*\Upsilon ^3+\text{Cosh}^4\rho (6n^2-6n)\nonumber\\
&+\text{Sinh}^4\rho
\left(6n^2+18n+12\right)+\text{Cosh}^2\rho \text{Sinh}^2\rho \left(36n^2+36n+12\right)\nonumber\\
&+\text{Cosh}^3\rho\text{Sinh$\rho $}\left(24n^2\right)+\text{Cosh$\rho
$} \text{Sinh}^3\rho  \left(24n^2+48n+24\right)\nonumber\\
&+\text{Cosh$\rho $} \text{Sinh$\rho $}(24n+12)\left\{1+2\Upsilon ^*\Upsilon-\Upsilon ^{*2}-\Upsilon
^2\right\} \nonumber\\
&+\text{Sinh}^2\rho (12n+12)\left\{1+2\Upsilon ^*\Upsilon -\Upsilon ^{*2}-\Upsilon ^2\right\}\nonumber\\
&+\text{Cosh}^2\rho (12n)\left\{1+2\Upsilon ^*\Upsilon
-\Upsilon ^{*2}-\Upsilon ^2\right\}\biggr].
\end{align}

Using Eqs. (\ref{17}-\ref{18}, \ref{28}-\ref{37}) the values of \(\left\langle :\overset{\wedge }\Pi ^2\overset{\wedge }\Phi ^2:\right\rangle {}_{\text{CSNS}}\) can be calculated as 

\begin{align} \label{69}
\left\langle :\overset{\wedge }\Pi ^2\overset{\wedge }\Phi ^2:\right\rangle _{\text{CSNS}}=&\frac{1}{4m^{2 }\mathbf{t}^2}\biggr[3+\Upsilon ^{*4}+\Upsilon ^4+6\Upsilon ^{*2}\Upsilon
^2+12\Upsilon ^*\Upsilon\nonumber\\
&-6\Upsilon ^{*2}-6\Upsilon ^2-4\Upsilon ^{*3}\Upsilon -4\Upsilon ^*\Upsilon ^3+\text{Cosh}^4\rho (6n^2-6n)\nonumber\\
&+\text{Sinh}^4\rho
\left(6n^2+18n+12\right)+\text{Cosh}^2\rho\text{Sinh}^2\rho
\left(36n^236n+12\right)\nonumber\\
&+\text{Cosh}^3\rho\text{Sinh$\rho $}\left(24n^2\right)\nonumber\\
&+\text{Cosh$\rho
$}\text{Sinh}^3\rho\left(24n^2+48n+24\right)\nonumber\\
&+\text{Cosh$\rho $}\text{Sinh$\rho $}(24n+12)\left\{1+2\Upsilon ^*\Upsilon -\Upsilon ^{*2}-\Upsilon
^2\right\}\nonumber\\
&+\text{Sinh}^2\rho (12n+12)\left\{1+2\Upsilon ^*\Upsilon -\Upsilon ^{*2}-\Upsilon ^2\right\}\nonumber\\
&+\text{Cosh}^2\rho (12n)\left\{1+2\Upsilon ^*\Upsilon
-\Upsilon ^{*2}-\Upsilon ^2\right\}\biggr].
\end{align}

Substituting the values of\(\left\langle :\overset{\wedge }\Phi ^4:\right\rangle _{\text{CSNS}}\),\(\left\langle :\overset{\wedge }\Pi ^4:\right\rangle _{\text{CSNS}}\),\(\left\langle
:\overset{\wedge }\Phi ^2\overset{\wedge }\Pi ^2:\right\rangle_{\text{CSNS}}\), \(\left\langle :\overset{\wedge }\Pi ^2\overset{\wedge }\Phi ^2:\right\rangle_{\text{CSNS}}\) in Eq. (\ref{57}), The Normal ordered value
of square of energy momentum tensor \(T_{00}\) can be calculated as 

\begin{align} \label{70}
\left\langle :T_{00}^2:\right\rangle _{\text{CSNS}}=&\left(\frac{1}{16m^{2}\mathbf{t}^4}+\frac{1}{8\mathbf{t}^2}+\frac{m^{2}}{16}\right)\biggr[3+\Upsilon
^{*4}+\Upsilon ^4\nonumber\\
&+6\Upsilon ^{*2}\Upsilon ^2+12\Upsilon ^*\Upsilon -6\Upsilon ^{*2}-6\Upsilon ^2-4\Upsilon ^{*3}\Upsilon -4\Upsilon ^*\Upsilon ^3\nonumber\\
&+\text{Cosh}^4\rho
(6n^2-6n)+\text{Sinh}^4\rho \left(6n^2+18n+12\right)\nonumber\\
&+\text{Cosh}^2\rho \text{Sinh}^2\rho \left(36n^2+36n+12\right)\nonumber\\
&+\text{Cosh}^3\rho \text{Sinh$\rho
$}\left(24n^2\right)+\text{Cosh$\rho $} \text{Sinh}^3\rho \left(24n^2+48n+24\right)\nonumber\\
&+\text{Cosh$\rho $} \text{Sinh$\rho $}(24n+12)\left\{1+2\Upsilon
^*\Upsilon -\Upsilon ^{*2}-\Upsilon ^2\right\}\nonumber\\
&+\text{Sinh}^2\rho (12n+12)\left\{1+2\Upsilon ^*\Upsilon -\Upsilon ^{*2}-\Upsilon ^2\right\}\nonumber\\
&+\text{Cosh}^2\rho(12n)\left\{1+2\Upsilon ^*\Upsilon
-\Upsilon ^{*2}-\Upsilon ^2\right\}\biggr].
\end{align}

Now, normal ordered value of temporal component of energy momentum tensor

\begin{equation} \label{71}
\left\langle :T_{00}:\right\rangle _{\text{CSNS}} =\frac{1}{2}\mathcal{G}^3 (\mathbf{t})m^{2 }\left\langle :\overset{\wedge }\Phi ^2:\right\rangle {}_{\text{CSNS}}+\frac{1}{2\mathcal{G}^3
(\mathbf{t})}\left\langle :\overset{\wedge }\Pi ^2:\right\rangle_{\text{CSNS}}.
\end{equation}

Using Eqs. (\ref{17}-\ref{18}, \ref{28}-\ref{37}) the values of \(\left\langle :\overset{\wedge }\Pi ^2:\right\rangle {}_{\text{CSNS}}\) can be calculated as

\begin{align} \label{74}
\left\langle :\overset{\wedge }\Pi ^2:\right\rangle {}_{\text{CSNS}}=&\frac{\mathcal{G}^3 (\mathbf{t})}{2m\mathbf{t}^2}[\text{Cosh}^2\rho (2n)+\text{Sinh}^2\rho
(2n+2)\nonumber\\
&+\text{Cosh$\rho $} \text{Sinh$\rho $}(4n+2)+1+\Upsilon ^*\Upsilon -\Upsilon ^{*2}-\Upsilon ^2],
\end{align}

Using Eqs. (\ref{17}-\ref{18}, \ref{28}-\ref{37}) the values of \(\left\langle :\overset{\wedge }\Phi ^2:\right\rangle {}_{\text{CSNS}}\) can be calculated as 

\begin{align} \label{77}
\left\langle :\overset{\wedge }\Phi ^2:\right\rangle_{\text{CSNS}}=&\frac{1}{2\text{m$\mathcal{G}$}^3 (\mathbf{t}}[\text{Cosh}^2\rho (2n)+\text{Sinh}^2\rho (2n+2)\nonumber\\
&+\text{Cosh$\rho
$} \text{Sinh$\rho $}(4n+2)+1+\Upsilon ^*\Upsilon -\Upsilon ^{*2}-\Upsilon ^2].
\end{align}

Using Eqs. (\ref{74}) and (\ref{77}) in Eqs. (\ref{71}), normal ordered average value of temporal component of energy momentum tensor can be computed as 

\begin{align} \label{78}
\left\langle :T_{00}:\right\rangle _{\text{CSNS}}=&\left[\frac{m}{4}+\frac{1}{4\text{m$\mathbf{t}$}^2}\right][\text{Cosh}^2\rho (2n)+\text{Sinh}^2\rho
(2n+2)\nonumber\\
&+\text{Cosh$\rho $} \text{Sinh$\rho $}(4n+2)+1+\Upsilon ^*\Upsilon -\Upsilon ^{*2}-\Upsilon ^2].
\end{align}

Square of of normal ordered value of temporal component of energy momentum tensor can be calculated as 

\begin{align} \label{79}
\left\langle :T_{00}:\rangle \right.{}^2{}_{\text{CSNS}}=&\left(\frac{1}{16m^{2 }\mathbf{t}^4}+\frac{1}{8\mathbf{t}^2}+\frac{m^{2 }}{16}\right)\biggr[1+\Upsilon
^{*4}+\Upsilon ^4+3\Upsilon ^{*2}\Upsilon ^2\nonumber\\
&+2\Upsilon ^*\Upsilon -2\Upsilon ^{*2}-2\Upsilon ^2-2\Upsilon ^{*3}\Upsilon -2\Upsilon ^*\Upsilon ^3\nonumber\\
&+\text{Cosh}^4\rho
\left(4n^2\right)+\text{Sinh}^4\rho \left(4n^2+8n+4\right)\nonumber\\
&+\text{Cosh}^2\rho\text{Sinh}^2\rho \left(24n^2+24n+4\right)+\text{Cosh}^3\rho \text{Sinh$\rho
$}(16n^2\nonumber\\
&+8n)+\text{Cosh$\rho $} \text{Sinh}^3\rho \left(16n^2+24n+8\right)\nonumber\\
&+\text{Cosh$\rho $} \text{Sinh$\rho $}(8n+4)\left\{1+\Upsilon
^*\Upsilon -\Upsilon ^{*2}-\Upsilon ^2\right\}\nonumber\\
&+\text{Sinh}^2\rho (4n+4)\left\{1+\Upsilon ^*\Upsilon -\Upsilon ^{*2}-\Upsilon ^2\right\}\nonumber\\
&+\text{Cosh}^2\rho
(4n)\left\{1+\Upsilon ^*\Upsilon -\Upsilon ^{*2}-\Upsilon ^2\right\}\biggr].
\end{align}

Substituting the values of \(\left\langle :T_{00}^2:\right\rangle {}_{\text{CSNS}}\) and \(\left\langle :T_{00}:\rangle \right.{}^2{}_{\text{CSNS}}\)
in Eq. (\ref{38}) then the density fluctuations for coherent squeezed number state is 

\begin{align} \label{80}
\triangle _{\text{CSNS}}=&\left(\frac{1}{16m^{2 }\mathbf{t}^4}+\frac{1}{8\mathbf{t}^2}+\frac{m^{2 }}{16}\right)\biggr[2+3\Upsilon ^{*2}\Upsilon ^2+10\Upsilon
^*\Upsilon -4\Upsilon ^{*2}\nonumber\\
&-4\Upsilon ^2-2\Upsilon ^{*3}\Upsilon -2\Upsilon ^*\Upsilon ^3+\text{Cosh}^4\rho (2n^2-6n)\nonumber\\
&+\text{Sinh}^4\rho \left(2n^2+10n+8\right)+\text{Cosh}^2\rho
 \text{Sinh}^2\rho \left(12n^2+12n+8\right)\nonumber\\
&+\text{Cosh}^3\rho\text{Sinh$\rho $}\left(8n^2-8n\right)+\text{Cosh$\rho $}\text{Sinh}^3\rho \left(8n^2+24n+16\right)\nonumber\\
&+\text{Cosh$\rho
$}\text{Sinh$\rho $}(16n+8)\left\{1-\varUpsilon ^{*2}-\varUpsilon ^2\right\}+\text{Cosh$\rho $} \text{Sinh$\rho $}(40n\nonumber\\
&+20)\left\{\varUpsilon ^*\varUpsilon
\right\}+\text{Sinh}^2\rho (8n+8)\left\{1-\varUpsilon^{*2}-\varUpsilon ^2\right\}\nonumber\\
&+\text{Sinh}^2\rho (20n+20)\left\{\varUpsilon ^*\varUpsilon \right\}+\text{Cosh}^2\rho
(8n)\left\{1-\varUpsilon ^{*2}-\varUpsilon ^2\right\}\nonumber\\
&+\text{Cosh}^2\rho (20n)\left\{\varUpsilon ^*\varUpsilon \right\}\biggr].   
\end{align}

\begin{figure}[h]%
\centering
\includegraphics[width=0.9\textwidth]{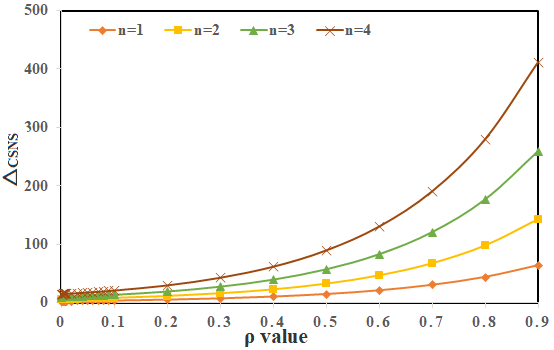}
\caption{ Variation of density fluctuations \(\triangle _{\text{CSNS}}\) for various values of squeezing parameter $\rho $}
\label{fig:figure_3}
\end{figure}

\begin{figure}[h]%
\centering
\includegraphics[width=0.9\textwidth]{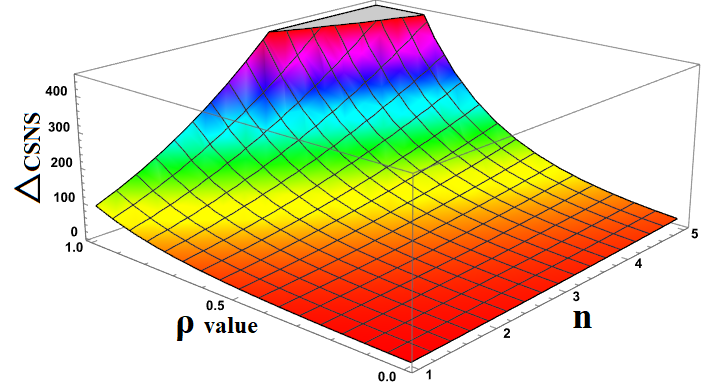}
\caption{3-D plot showing variation of density fluctuations \(\triangle _{\text{CSNS}}\) with number state and squeezing parameter
$\rho $}
\label{fig:figure_4}
\end{figure}

Eq. (\ref{80}) shows the \(\triangle _{\text{CSNS}}\) (density fluctuations for Coherent Squeezed Number State (CSNS)) that depend on squeezing parameter
($\rho $), Coherent state parameter ($\varUpsilon $) as well as number state (n) but its dependency on Coherent
state parameter and number state of consideration is more prominent than the squeezing parameter $\rho$. Numerical values of density fluctuations
\(\triangle _{\text{CSNS}}\) for various values ofcoherent squeezed number state calculated using Eq. (\ref{80}) for n =1 to 4 is as shown in Table  \ref{tab:table_4}, \ref{tab:table_5},  \ref{tab:table_6}. For simplification we considered, $\varUpsilon
$*=$\varUpsilon $=1, and for squeezing parameter ($\rho $) ranging between 0 .001
to 0.09 the variation in density fluctuations is very small for all number state taken into consideration in Table \ref{tab:table_4},  \ref{tab:table_5} but in Table \ref{tab:table_6} it is shown that for
$\rho $ ranging 0.1 to 0.09 shows large variation in \(\triangle _{\text{CSNS}}\). If we consider coherent state parameter $\varUpsilon $*=$\varUpsilon
$=0, Eq. (\ref{80}) reduces to Eq. (\ref{56}) that is for density fluctuations for squeezed number state. For substituting both coherent state parameter
$\varUpsilon $*=$\varUpsilon $=0 and number state n=0 in Eq. (\ref{80}), it produces density fluctuations for squeezed vacuum state \cite{venkataratnam_density_2008, venkataratnam_oscillatory_2010, venkataratnam_behavior_2013}, showing the
validity of approximation used in SCEE. Fig.  \ref{fig:figure_3} shows variation of density fluctuations \(\triangle _{\text{CSNS}}\), for various values of coherent squeezed
number state with squeezing parameter $\rho $. It can be deduced from Fig.  \ref{fig:figure_3} that with increasing value of $\rho $ as well as n, density fluctuations
\(\triangle _{\text{CSNS}}\) increases. 3-D plot between $\rho $, n and \(\triangle _{\text{CSNS}}\), shows variation of density fluctuations \(\triangle
_{\text{CSNS}}\) with various values of coherent squeezed number state and squeezing parameter $\rho $ (Fig. \ref{fig:figure_4}).

\begin{table}[h]\label{table 4}
\caption{Numerical values of density fluctuations \(\triangle _{\text{CSNS}}\) for various { }coherent squeezed number state for squeezing parameter
$\rho <<<$1}\label{t4}%
\label{tab:table_4}
\begin{tabular}{@{}llllll@{}}
\toprule
$\rho$ & n=1  & n=2 & n=3 & n=4\\
\midrule
0.001 & 2.75902 & 5.76903 & 9.78306 & 14.80110 \\
 0.002 & 2.76806 & 5.78813 & 9.81624 & 14.85240 \\
 0.003 & 2.77714 & 5.80730 & 9.84953 & 14.90380 \\
 0.004 & 2.78624 & 5.82653 & 9.88295 & 14.95550 \\
 0.005 & 2.79538 & 5.84583 & 9.91648 & 15.00730 \\
 0.006 & 2.80454 & 5.86520 & 9.95014 & 15.05940 \\
 0.007 & 2.81374 & 5.88463 & 9.98392 & 15.11160 \\
 0.008 & 2.82297 & 5.90413 & 10.01780 & 15.16400 \\
 0.009 & 2.83223 & 5.92370 & 10.05180 & 15.21660 \\
\botrule
\end{tabular}
\end{table}

\begin{table}[h]\label{table 5}
\caption{Numerical values of density fluctuations \(\triangle _{\text{CSNS}}\) for various { }coherent squeezed number state for squeezing parameter
$\rho <<$1}\label{t5}%
\label{tab:table_5}
\begin{tabular}{@{}llllll@{}}
\toprule
 $\rho$  & n=1 & n=2 & n=3 & n=4 \\
\midrule
 0.01 & 2.84152 & 5.94334 & 10.08600 & 15.26940 \\
 0.02 & 2.93615 & 6.14353 & 10.43420 & 15.80820 \\
 0.03 & 3.03400 & 6.35083 & 10.79520 & 16.36700 \\
 0.04 & 3.13520 & 6.56551 & 11.16930 & 16.94670 \\
 0.05 & 3.23986 & 6.78784 & 11.55720 & 17.54800 \\
 0.06 & 3.34812 & 7.01811 & 11.95940 & 18.17190 \\
 0.07 & 3.46011 & 7.25664 & 12.37630 & 18.81910 \\
 0.08 & 3.57596 & 7.50372 & 12.80860 & 19.49060 \\
 0.09 & 3.69582 & 7.75970 & 13.25690 & 20.18740 \\
\botrule
\end{tabular}
\end{table}

\begin{table}[h]
\caption{Numerical values of density fluctuations \(\triangle _{\text{CSNS}}\) for various { }coherent squeezed number state for squeezing parameter
$\rho <$1}\label{t6}%
\label{tab:table_6}
\begin{tabular}{@{}llllll@{}}
\toprule
$\rho$  & n=1  & n=2 & n=3 & n=4\\
\midrule
 0.1 & 3.81984 & 8.02489 & 13.72180 & 20.91050 \\
 0.2 & 5.32605 & 11.26900 & 19.43740 & 29.83140 \\
 0.3 & 7.46335 & 15.92570 & 27.70820 & 42.81080 \\
 0.4 & 10.51790 & 22.64950 & 39.73410 & 61.77180 \\
 0.5 & 14.91100 & 32.40740 & 57.29290 & 89.56740 \\
 0.6 & 21.26490 & 46.63140 & 83.02110 & 130.43400 \\
 0.7 & 30.49980 & 67.44430 & 120.83300 & 190.66700 \\
 0.8 & 43.97830 & 97.99640 & 176.54700 & 279.63000 \\
 0.9 & 63.72180 & 142.96800 & 258.81200 & 411.2550\\
\botrule
\end{tabular}
\end{table}

\section{Results and Discussion}

We analyzed quantum effect of Squeezed Number State and Coherent Squeezed Number State on Friedmann-Robertson-Walker universe in the framework of semiclassical theory of gravity. Non-classical state of gravity plays a significant role in number state representation and particle production. As there was no consistent quantum theory of gravity is available, so semiclassical methods at appropriate situations, where background metric has classical consideration with matter field as  quantum effects \cite{anderson_effects_1983, campos_semiclassical_1994}. \\
Here, we consider number state evolution for oscillatory phase inflaton for Coherent Squeezed Number State and Squeezed Number State formalisms were analyzed. We analyzed quantum consideration for coherent state depends on number state and coherent state parameter while squeezed states are concomitant on number state
and parameter of squeezing. Both these formulations obtained universe at squeezed vacuum state \cite{venkataratnam_density_2008} shows the validity of the semiclassical
Einstein equation and energy density of the inflaton. Oscillatory phase of inflaton for Coherent Squeezed Number State and Squeezed Number State has similar power-law expansion of Einstein equation \cite{lachieze-rey_cosmological_1999,takahashi_thermo_1996, xu_quantum_2007}. In our study, the results revealed that for both Squeezed number states and
Coherent squeezed number states and density fluctuations are inversely proportional to various powers to t.\\
The results deduced that inflaton is minimally coupled to flat FRW universe can be represented by SCEE. Quantum effects and density fluctuations
of inflaton are analyzed in context of number state representation for Squeezed Number
State (SNS) and Coherent Squeezed Number State (CSNS), further generalized to demonstrate the validity of approximations in terms of energy density. The semiclassical theory successfully demonstrates that production of particles due to the quantum behavior of number state of consideration for
Squeezed Number State (SNS) and Coherent Squeezed Number State (CSNS) in terms of a massive inflaton.\\
Fig. \ref{fig:figure_1},  \ref{fig:figure_3} plotted for density fluctuations in Squeezed Number State (SNS) and Coherent Squeezed Number State (CSNS) respectively shows increase in density fluctuations with increasing parameter of squeezing ($\rho$). 3-D plot between $\rho $, n, \(\triangle\) shows variation of density fluctuations in Squeezed Number State (SNS) (Fig. \ref{fig:figure_2}) and Coherent Squeezed Number State (CSNS) (Fig.  \ref{fig:figure_4}) respectively. Here we observed
that density fluctuations increase for both increasing values of number state as well as parameter of squeezing ($\rho$). \\

\bibliography{sn-bibliography.bib}

\begin{thebibliography}{10}
\expandafter\ifx\csname url\endcsname\relax
  \def\url#1{\burl{#1}}\fi
\expandafter\ifx\csname urlprefix\endcsname\relax\def\urlprefix{URL }\fi
\providecommand{\bibinfo}[2]{#2}
\providecommand{\eprint}[2][]{\url{#2}}
\providecommand{\doi}[1]{\url{https://doi.org/#1}}
\bibcommenthead

\bibitem{kubik_origin_2022}
\bibinfo{author}{Kubik, B.}, \bibinfo{author}{Karska, A.} \& \bibinfo{author}{Opitom, C.}
\newblock \bibinfo{title}{Origin of the universe and planetary systems}  (\bibinfo{year}{2022}).
\newblock \urlprefix\url{https://books.rsc.org/books/edited-volume/2003/chapter/4583607/Origin-of-the-Universe-and-Planetary-Systems}.

\bibitem{moore_big_2014}
\bibinfo{author}{Moore, B.} \& \bibinfo{author}{Moore, B.}
\newblock \bibinfo{title}{The big bang}.
\newblock \emph{\bibinfo{journal}{Elephants in Space: The Past, Present and Future of Life and the Universe}} \bibinfo{pages}{43--55} (\bibinfo{year}{2014}).
\newblock \urlprefix\url{https://link.springer.com/chapter/10.1007/978-3-319-05672-2_3}.

\bibitem{green_cosmological_2022}
\bibinfo{author}{Green, D.}, \bibinfo{author}{Guo, Y.} \& \bibinfo{author}{Wallisch, B.}
\newblock \bibinfo{title}{Cosmological implications of axion-matter couplings}.
\newblock \emph{\bibinfo{journal}{Journal of Cosmology and Astroparticle Physics}} \textbf{\bibinfo{volume}{2022}}, \bibinfo{pages}{019} (\bibinfo{year}{2022}).
\newblock \urlprefix\url{https://iopscience.iop.org/article/10.1088/1475-7516/2022/02/019/meta}.

\bibitem{albrecht_inflation_1994}
\bibinfo{author}{Albrecht, A.}, \bibinfo{author}{Ferreira, P.}, \bibinfo{author}{Joyce, M.} \& \bibinfo{author}{Prokopec, T.}
\newblock \bibinfo{title}{Inflation and squeezed quantum states}.
\newblock \emph{\bibinfo{journal}{Physical Review D}} \textbf{\bibinfo{volume}{50}}, \bibinfo{pages}{4807--4820} (\bibinfo{year}{1994}).
\newblock \urlprefix\url{https://link.aps.org/doi/10.1103/PhysRevD.50.4807}.

\bibitem{albrecht_reheating_1982}
\bibinfo{author}{Albrecht, A.}, \bibinfo{author}{Steinhardt, P.~J.}, \bibinfo{author}{Turner, M.~S.} \& \bibinfo{author}{Wilczek, F.}
\newblock \bibinfo{title}{Reheating an {Inflationary} {Universe}}.
\newblock \emph{\bibinfo{journal}{Physical Review Letters}} \textbf{\bibinfo{volume}{48}}, \bibinfo{pages}{1437--1440} (\bibinfo{year}{1982}).

\bibitem{kofman_reheating_1994}
\bibinfo{author}{Kofman, L.}, \bibinfo{author}{Linde, A.} \& \bibinfo{author}{Starobinsky, A.~A.}
\newblock \bibinfo{title}{Reheating after {Inflation}}.
\newblock \emph{\bibinfo{journal}{Physical Review Letters}} \textbf{\bibinfo{volume}{73}}, \bibinfo{pages}{3195--3198} (\bibinfo{year}{1994}).

\bibitem{allahverdi_reheating_2010}
\bibinfo{author}{Allahverdi, R.}, \bibinfo{author}{Brandenberger, R.}, \bibinfo{author}{Cyr-Racine, F.-Y.} \& \bibinfo{author}{Mazumdar, A.}
\newblock \bibinfo{title}{Reheating in {Inflationary} {Cosmology}: {Theory} and {Applications}}.
\newblock \emph{\bibinfo{journal}{Annual Review of Nuclear and Particle Science}} \textbf{\bibinfo{volume}{60}}, \bibinfo{pages}{27--51} (\bibinfo{year}{2010}).

\bibitem{cook_reheating_2015}
\bibinfo{author}{Cook, J.~L.}, \bibinfo{author}{Dimastrogiovanni, E.}, \bibinfo{author}{Easson, D.~A.} \& \bibinfo{author}{Krauss, L.~M.}
\newblock \bibinfo{title}{Reheating predictions in single field inflation}.
\newblock \emph{\bibinfo{journal}{Journal of Cosmology and Astroparticle Physics}} \textbf{\bibinfo{volume}{2015}}, \bibinfo{pages}{047} (\bibinfo{year}{2015}).
\newblock \urlprefix\url{https://dx.doi.org/10.1088/1475-7516/2015/04/047}.

\bibitem{yadav_reheating_2023}
\bibinfo{author}{Yadav, S.}, \bibinfo{author}{Goswami, R.}, \bibinfo{author}{Venkataratnam, K.~K.} \& \bibinfo{author}{Yajnik, U.~A.}
\newblock \bibinfo{title}{Reheating constraints on modified quadratic chaotic inflation}.
\newblock \urlprefix\url{http://arxiv.org/abs/2309.06990}.
\newblock \eprint{2309.06990 [astro-ph]}.

\bibitem{yadav2024reheating}
\bibinfo{author}{Yadav, S.}, \bibinfo{author}{Gangal, D.} \& \bibinfo{author}{Venkataratnam, K.~K.}
\newblock \bibinfo{title}{Reheating constraints on mutated hilltop inflation} (\bibinfo{year}{2024}).
\newblock \eprint{2401.09806}.

\bibitem{mohajan_friedmann_2013}
\bibinfo{author}{Mohajan, H.}
\newblock \bibinfo{title}{Friedmann, {Robertson}-{Walker} ({FRW}) {Models} in {Cosmology}} (\bibinfo{year}{2013}).
\newblock \urlprefix\url{https://mpra.ub.uni-muenchen.de/52402/}.

\bibitem{suresh_particle_2004}
\bibinfo{author}{Suresh, P.~K.}
\newblock \bibinfo{title}{Particle {Creation} in the {Oscillatory} {Phase} of {Inflaton}}.
\newblock \emph{\bibinfo{journal}{International Journal of Theoretical Physics}} \textbf{\bibinfo{volume}{43}}, \bibinfo{pages}{425--436} (\bibinfo{year}{2004}).
\newblock \urlprefix\url{http://link.springer.com/10.1023/B:IJTP.0000028875.07382.4e}.

\bibitem{kim_one-parameter_1999}
\bibinfo{author}{Kim, J.~K.} \& \bibinfo{author}{Kim, S.~P.}
\newblock \bibinfo{title}{One-parameter squeezed {Gaussian} states of a time-dependent harmonic oscillator and the selection rule for vacuum states}.
\newblock \emph{\bibinfo{journal}{Journal of Physics A: Mathematical and General}} \textbf{\bibinfo{volume}{32}}, \bibinfo{pages}{2711--2718} (\bibinfo{year}{1999}).
\newblock \urlprefix\url{https://iopscience.iop.org/article/10.1088/0305-4470/32/14/012}.

\bibitem{finelli_quantum_1999}
\bibinfo{author}{Finelli, F.}, \bibinfo{author}{Gruppuso, A.} \& \bibinfo{author}{Venturi, G.}
\newblock \bibinfo{title}{Quantum fields in an expanding universe}.
\newblock \emph{\bibinfo{journal}{Classical and Quantum Gravity}} \textbf{\bibinfo{volume}{16}}, \bibinfo{pages}{3923--3935} (\bibinfo{year}{1999}).
\newblock \urlprefix\url{https://iopscience.iop.org/article/10.1088/0264-9381/16/12/310}.

\bibitem{geralico_novel_2004}
\bibinfo{author}{Geralico, A.}, \bibinfo{author}{Landolfi, G.}, \bibinfo{author}{Ruggeri, G.} \& \bibinfo{author}{Soliani, G.}
\newblock \bibinfo{title}{Novel approach to the study of quantum effects in the early {Universe}}.
\newblock \emph{\bibinfo{journal}{Physical Review D}} \textbf{\bibinfo{volume}{69}}, \bibinfo{pages}{043504} (\bibinfo{year}{2004}).
\newblock \urlprefix\url{https://link.aps.org/doi/10.1103/PhysRevD.69.043504}.

\bibitem{padmanabhan_gravity_2005}
\bibinfo{author}{Padmanabhan, T.}
\newblock \bibinfo{title}{Gravity and the thermodynamics of horizons}.
\newblock \emph{\bibinfo{journal}{Physics Reports}} \textbf{\bibinfo{volume}{406}}, \bibinfo{pages}{49--125} (\bibinfo{year}{2005}).
\newblock \urlprefix\url{https://linkinghub.elsevier.com/retrieve/pii/S0370157304004582}.

\bibitem{mahajan_particle_2008}
\bibinfo{author}{Mahajan, G.} \& \bibinfo{author}{Padmanabhan, T.}
\newblock \bibinfo{title}{Particle creation, classicality and related issues in quantum field theory: {I}. {Formalism} and toy models}.
\newblock \emph{\bibinfo{journal}{General Relativity and Gravitation}} \textbf{\bibinfo{volume}{40}}, \bibinfo{pages}{661--708} (\bibinfo{year}{2008}).
\newblock \urlprefix\url{http://link.springer.com/10.1007/s10714-007-0526-z}.

\bibitem{lachieze-rey_cosmic_1995}
\bibinfo{author}{Lachièze-Rey, M.} \& \bibinfo{author}{Luminet, J.-P.}
\newblock \bibinfo{title}{Cosmic topology}.
\newblock \emph{\bibinfo{journal}{Physics Reports}} \textbf{\bibinfo{volume}{254}}, \bibinfo{pages}{135--214} (\bibinfo{year}{1995}).
\newblock \urlprefix\url{https://linkinghub.elsevier.com/retrieve/pii/037015739400085H}.

\bibitem{ellis_cosmological_1998}
\bibinfo{author}{Ellis, G. F.~R.} \& \bibinfo{author}{van Elst, H.}
\newblock \bibinfo{title}{Cosmological models ({Cargèse} lectures 1998)}  (\bibinfo{year}{1998}).
\newblock \urlprefix\url{https://arxiv.org/abs/gr-qc/9812046}.

\bibitem{carvalho_scalar_2004}
\bibinfo{author}{Carvalho, A. D.~M.}, \bibinfo{author}{Furtado, C.} \& \bibinfo{author}{Pedrosa, I.~A.}
\newblock \bibinfo{title}{Scalar fields and exact invariants in a {Friedmann}-{Robertson}-{Walker} spacetime}.
\newblock \emph{\bibinfo{journal}{Physical Review D}} \textbf{\bibinfo{volume}{70}}, \bibinfo{pages}{123523} (\bibinfo{year}{2004}).
\newblock \urlprefix\url{https://link.aps.org/doi/10.1103/PhysRevD.70.123523}.

\bibitem{bakke_geometric_2009}
\bibinfo{author}{Bakke, K.}, \bibinfo{author}{Pedrosa, I.~A.} \& \bibinfo{author}{Furtado, C.}
\newblock \bibinfo{title}{Geometric phases and squeezed quantum states of relic gravitons}.
\newblock \emph{\bibinfo{journal}{Journal of Mathematical Physics}} \textbf{\bibinfo{volume}{50}}, \bibinfo{pages}{113521} (\bibinfo{year}{2009}).
\newblock \urlprefix\url{https://pubs.aip.org/aip/jmp/article/930980}.

\bibitem{kennard_zur_1927}
\bibinfo{author}{Kennard, E.~H.}
\newblock \bibinfo{title}{Zur {Quantenmechanik} einfacher {Bewegungstypen}}.
\newblock \emph{\bibinfo{journal}{Zeitschrift for Physik}} \textbf{\bibinfo{volume}{44}}, \bibinfo{pages}{326--352} (\bibinfo{year}{1927}).
\newblock \urlprefix\url{http://link.springer.com/10.1007/BF01391200}.

\bibitem{venkataratnam_particle_2004}
\bibinfo{author}{Venkataratnam, K.~K.} \& \bibinfo{author}{Suresh, P.~K.}
\newblock \bibinfo{title}{{PARTICLE} {PRODUCTION} {OF} {COHERENTLY} {OSCILLATING} {NONCLASSICAL} {INFLATON} {IN} {FRW} {UNIVERSE}}.
\newblock \emph{\bibinfo{journal}{International Journal of Modern Physics D}} \textbf{\bibinfo{volume}{13}}, \bibinfo{pages}{239--252} (\bibinfo{year}{2004}).
\newblock \urlprefix\url{https://www.worldscientific.com/doi/abs/10.1142/S0218271804004578}.

\bibitem{pedrosa_gaussian_2009}
\bibinfo{author}{Pedrosa, I.}, \bibinfo{author}{Bakke, K.} \& \bibinfo{author}{Furtado, C.}
\newblock \bibinfo{title}{Gaussian wave packet states of relic gravitons}.
\newblock \emph{\bibinfo{journal}{Physics Letters B}} \textbf{\bibinfo{volume}{671}}, \bibinfo{pages}{314--317} (\bibinfo{year}{2009}).
\newblock \urlprefix\url{https://linkinghub.elsevier.com/retrieve/pii/S0370269308014822}.

\bibitem{stoica_friedmann-lemaitre-robertson-walker_2016}
\bibinfo{author}{Stoica, O.~C.}
\newblock \bibinfo{title}{The {Friedmann}-{Lemaître}-{Robertson}-{Walker} {Big} {Bang} {Singularities} are {Well} {Behaved}}.
\newblock \emph{\bibinfo{journal}{International Journal of Theoretical Physics}} \textbf{\bibinfo{volume}{55}}, \bibinfo{pages}{71--80} (\bibinfo{year}{2016}).
\newblock \urlprefix\url{http://link.springer.com/10.1007/s10773-015-2634-y}.

\bibitem{hu_anisotropy_1978}
\bibinfo{author}{Hu, B.~L.} \& \bibinfo{author}{Parker, L.}
\newblock \bibinfo{title}{Anisotropy damping through quantum effects in the early universe}.
\newblock \emph{\bibinfo{journal}{Physical Review D}} \textbf{\bibinfo{volume}{17}}, \bibinfo{pages}{933--945} (\bibinfo{year}{1978}).
\newblock \urlprefix\url{https://link.aps.org/doi/10.1103/PhysRevD.17.933}.

\bibitem{dhayal_quantum_2020}
\bibinfo{author}{Dhayal, R.}, \bibinfo{author}{Rathore, M.} \& \bibinfo{author}{Venkataratnam, K.~K.}
\newblock \bibinfo{title}{Quantum fluctuations and particle production in the oscillatory phase of a thermal inflaton in a {FRW} universe}.
\newblock \emph{\bibinfo{journal}{Modern Physics Letters A}} \textbf{\bibinfo{volume}{35}}, \bibinfo{pages}{2050022} (\bibinfo{year}{2020}).
\newblock \urlprefix\url{https://www.worldscientific.com/doi/abs/10.1142/S0217732320500224}.

\bibitem{fischetti_quantum_1979}
\bibinfo{author}{Fischetti, M.~V.}, \bibinfo{author}{Hartle, J.~B.} \& \bibinfo{author}{Hu, B.~L.}
\newblock \bibinfo{title}{Quantum effects in the early universe. {I}. {Influence} of trace anomalies on homogeneous, isotropic, classical geometries}.
\newblock \emph{\bibinfo{journal}{Physical Review D}} \textbf{\bibinfo{volume}{20}}, \bibinfo{pages}{1757--1771} (\bibinfo{year}{1979}).
\newblock \urlprefix\url{https://link.aps.org/doi/10.1103/PhysRevD.20.1757}.

\bibitem{hartle_quantum_1979}
\bibinfo{author}{Hartle, J.~B.} \& \bibinfo{author}{Hu, B.~L.}
\newblock \bibinfo{title}{Quantum effects in the early universe. {II}. {Effective} action for scalar fields in homogeneous cosmologies with small anisotropy}.
\newblock \emph{\bibinfo{journal}{Physical Review D}} \textbf{\bibinfo{volume}{20}}, \bibinfo{pages}{1772--1782} (\bibinfo{year}{1979}).
\newblock \urlprefix\url{https://link.aps.org/doi/10.1103/PhysRevD.20.1772}.

\bibitem{hartle_quantum_1980}
\bibinfo{author}{Hartle, J.~B.} \& \bibinfo{author}{Hu, B.~L.}
\newblock \bibinfo{title}{Quantum effects in the early universe. {III}. {Dissipation} of anisotropy by scalar particle production}.
\newblock \emph{\bibinfo{journal}{Physical Review D}} \textbf{\bibinfo{volume}{21}}, \bibinfo{pages}{2756--2769} (\bibinfo{year}{1980}).
\newblock \urlprefix\url{https://link.aps.org/doi/10.1103/PhysRevD.21.2756}.

\bibitem{hartle_quantum_1981}
\bibinfo{author}{Hartle, J.~B.}
\newblock \bibinfo{title}{Quantum effects in the early universe. {V}. {Finite} particle production without trace anomalies}.
\newblock \emph{\bibinfo{journal}{Physical Review D}} \textbf{\bibinfo{volume}{23}}, \bibinfo{pages}{2121--2128} (\bibinfo{year}{1981}).
\newblock \urlprefix\url{https://link.aps.org/doi/10.1103/PhysRevD.23.2121}.

\bibitem{anderson_effects_1983}
\bibinfo{author}{Anderson, P.}
\newblock \bibinfo{title}{Effects of quantum fields on singularities and particle horizons in the early universe}.
\newblock \emph{\bibinfo{journal}{Physical Review D}} \textbf{\bibinfo{volume}{28}}, \bibinfo{pages}{271--285} (\bibinfo{year}{1983}).
\newblock \urlprefix\url{https://link.aps.org/doi/10.1103/PhysRevD.28.271}.

\bibitem{campos_semiclassical_1994}
\bibinfo{author}{Campos, A.} \& \bibinfo{author}{Verdaguer, E.}
\newblock \bibinfo{title}{Semiclassical equations for weakly inhomogeneous cosmologies}.
\newblock \emph{\bibinfo{journal}{Physical Review D}} \textbf{\bibinfo{volume}{49}}, \bibinfo{pages}{1861--1880} (\bibinfo{year}{1994}).
\newblock \urlprefix\url{https://link.aps.org/doi/10.1103/PhysRevD.49.1861}.

\bibitem{geralico_novel_2004-1}
\bibinfo{author}{Geralico, A.}, \bibinfo{author}{Landolfi, G.}, \bibinfo{author}{Ruggeri, G.} \& \bibinfo{author}{Soliani, G.}
\newblock \bibinfo{title}{Novel approach to the study of quantum effects in the early {Universe}}.
\newblock \emph{\bibinfo{journal}{Physical Review D}} \textbf{\bibinfo{volume}{69}}, \bibinfo{pages}{043504} (\bibinfo{year}{2004}).
\newblock \urlprefix\url{https://link.aps.org/doi/10.1103/PhysRevD.69.043504}.

\bibitem{pedrosa_exact_2007}
\bibinfo{author}{Pedrosa, I.}, \bibinfo{author}{Furtado, C.} \& \bibinfo{author}{Rosas, A.}
\newblock \bibinfo{title}{Exact linear invariants and quantum effects in the early universe}.
\newblock \emph{\bibinfo{journal}{Physics Letters B}} \textbf{\bibinfo{volume}{651}}, \bibinfo{pages}{384--387} (\bibinfo{year}{2007}).
\newblock \urlprefix\url{https://linkinghub.elsevier.com/retrieve/pii/S0370269307007599}.

\bibitem{lopes_gaussian_2009}
\bibinfo{author}{Lopes, C. E.~F.}, \bibinfo{author}{Pedrosa, I.~A.}, \bibinfo{author}{Furtado, C.} \& \bibinfo{author}{De~M.~Carvalho, A.~M.}
\newblock \bibinfo{title}{Gaussian wave packet states of scalar fields in a universe of de {Sitter}}.
\newblock \emph{\bibinfo{journal}{Journal of Mathematical Physics}} \textbf{\bibinfo{volume}{50}}, \bibinfo{pages}{083511} (\bibinfo{year}{2009}).
\newblock \urlprefix\url{https://pubs.aip.org/aip/jmp/article/920134}.

\bibitem{kuo_semiclassical_1993}
\bibinfo{author}{Kuo, C.-I.} \& \bibinfo{author}{Ford, L.~H.}
\newblock \bibinfo{title}{Semiclassical gravity theory and quantum fluctuations}.
\newblock \emph{\bibinfo{journal}{Physical Review D}} \textbf{\bibinfo{volume}{47}}, \bibinfo{pages}{4510--4519} (\bibinfo{year}{1993}).
\newblock \urlprefix\url{https://link.aps.org/doi/10.1103/PhysRevD.47.4510}.

\bibitem{caves_quantum-mechanical_1981}
\bibinfo{author}{Caves, C.~M.}
\newblock \bibinfo{title}{Quantum-mechanical noise in an interferometer}.
\newblock \emph{\bibinfo{journal}{Physical Review D}} \textbf{\bibinfo{volume}{23}}, \bibinfo{pages}{1693--1708} (\bibinfo{year}{1981}).
\newblock \urlprefix\url{https://link.aps.org/doi/10.1103/PhysRevD.23.1693}.

\bibitem{matacz_coherent_1994}
\bibinfo{author}{Matacz, A.~L.}
\newblock \bibinfo{title}{Coherent state representation of quantum fluctuations in the early {Universe}}.
\newblock \emph{\bibinfo{journal}{Physical Review D}} \textbf{\bibinfo{volume}{49}}, \bibinfo{pages}{788--798} (\bibinfo{year}{1994}).
\newblock \urlprefix\url{https://link.aps.org/doi/10.1103/PhysRevD.49.788}.

\bibitem{suresh_thermal_2001}
\bibinfo{author}{Suresh, P.~K.}
\newblock \bibinfo{title}{{THERMAL} {SQUEEZING} {AND} {DENSITY} {FLUCTUATIONS} {IN} {SEMICLASSICAL} {THEORY} {OF} {GRAVITY}}.
\newblock \emph{\bibinfo{journal}{Modern Physics Letters A}} \textbf{\bibinfo{volume}{16}}, \bibinfo{pages}{707--717} (\bibinfo{year}{2001}).
\newblock \urlprefix\url{https://www.worldscientific.com/doi/abs/10.1142/S0217732301003802}.

\bibitem{suresh_squeezed_1998}
\bibinfo{author}{Suresh, P.~K.} \& \bibinfo{author}{Kuriakose, V.~C.}
\newblock \bibinfo{title}{{SQUEEZED} {STATE} {REPRESENTATION} {OF} {QUANTUM} {FLUCTUATIONS} {AND} {SEMICLASSICAL} {THEORY}}.
\newblock \emph{\bibinfo{journal}{Modern Physics Letters A}} \textbf{\bibinfo{volume}{13}}, \bibinfo{pages}{165--172} (\bibinfo{year}{1998}).
\newblock \urlprefix\url{https://www.worldscientific.com/doi/abs/10.1142/S0217732398000218}.

\bibitem{suresh_nonclassical_2001}
\bibinfo{author}{Suresh, P.~K.}
\newblock \bibinfo{title}{{NONCLASSICAL} {STATE} {REPRESENTATION} {OF} {INFLATON} {AND} {POWER}-{LAW} {EXPANSION} {IN} {FRW} {UNIVERSE}}.
\newblock \emph{\bibinfo{journal}{Modern Physics Letters A}} \textbf{\bibinfo{volume}{16}}, \bibinfo{pages}{2431--2438} (\bibinfo{year}{2001}).
\newblock \urlprefix\url{https://www.worldscientific.com/doi/abs/10.1142/S0217732301005874}.

\bibitem{venkataratnam_nonclassical_2010}
\bibinfo{author}{Venkataratnam, K.~K.} \& \bibinfo{author}{Suresh, P.~K.}
\newblock \bibinfo{title}{{NONCLASSICAL} {SCALAR} {FIELD} {IN} {THE} {FRW} {UNIVERSE}}.
\newblock \emph{\bibinfo{journal}{International Journal of Modern Physics D}} \textbf{\bibinfo{volume}{19}}, \bibinfo{pages}{37--61} (\bibinfo{year}{2010}).
\newblock \urlprefix\url{https://www.worldscientific.com/doi/abs/10.1142/S021827181001621X}.

\bibitem{venkataratnam_density_2008}
\bibinfo{author}{Venkataratnam, K.~K.} \& \bibinfo{author}{Suresh, P.~K.}
\newblock \bibinfo{title}{{DENSITY} {FLUCTUATIONS} {IN} {THE} {OSCILLATORY} {PHASE} {OF} {A} {NONCLASSICAL} {INFLATON} {IN} {THE} {FRW} {UNIVERSE}}.
\newblock \emph{\bibinfo{journal}{International Journal of Modern Physics D}} \textbf{\bibinfo{volume}{17}}, \bibinfo{pages}{1991--2005} (\bibinfo{year}{2008}).
\newblock \urlprefix\url{https://www.worldscientific.com/doi/abs/10.1142/S0218271808013662}.

\bibitem{venkataratnam_oscillatory_2010}
\bibinfo{author}{Venkataratnam, K.~K.} \& \bibinfo{author}{Suresh, P.~K.}
\newblock \bibinfo{title}{{OSCILLATORY} {PHASE} {OF} {NONCLASSICAL} {THERMAL} {INFLATON} {IN} {FRW} {UNIVERSE}}.
\newblock \emph{\bibinfo{journal}{International Journal of Modern Physics D}} \textbf{\bibinfo{volume}{19}}, \bibinfo{pages}{1147--1195} (\bibinfo{year}{2010}).
\newblock \urlprefix\url{https://www.worldscientific.com/doi/abs/10.1142/S0218271810017184}.

\bibitem{venkataratnam_behavior_2013}
\bibinfo{author}{Venkataratnam, K.~K.}
\newblock \bibinfo{title}{{BEHAVIOR} {OF} {NON}-{CLASSICAL} {INFLATON} {IN} {THE} {FRW} {UNIVERSE}}.
\newblock \emph{\bibinfo{journal}{Modern Physics Letters A}} \textbf{\bibinfo{volume}{28}}, \bibinfo{pages}{1350168} (\bibinfo{year}{2013}).
\newblock \urlprefix\url{https://www.worldscientific.com/doi/abs/10.1142/S021773231350168X}.

\bibitem{dhayal_quantum_2020-1}
\bibinfo{author}{Dhayal, R.}, \bibinfo{author}{Rathore, M.} \& \bibinfo{author}{Venkataratnam, K.~K.}
\newblock \bibinfo{title}{Quantum fluctuations and particle production in the oscillatory phase of a thermal inflaton in a {FRW} universe}.
\newblock \emph{\bibinfo{journal}{Modern Physics Letters A}} \textbf{\bibinfo{volume}{35}}, \bibinfo{pages}{2050022} (\bibinfo{year}{2020}).
\newblock \urlprefix\url{https://www.worldscientific.com/doi/abs/10.1142/S0217732320500224}.

\bibitem{lachieze-rey_cosmological_1999}
\bibinfo{author}{Ellis, G. F.~R.} \& \bibinfo{author}{Elst, H.}
\newblock \bibinfo{title}{ in \textit{Cosmological {Models}}} (ed.\bibinfo{editor}{Lachièze-Rey, M.}) \emph{\bibinfo{booktitle}{Theoretical and {Observational} {Cosmology}}} \bibinfo{pages}{1--116} (\bibinfo{publisher}{Springer Netherlands}, \bibinfo{address}{Dordrecht}, \bibinfo{year}{1999}).
\newblock \urlprefix\url{http://link.springer.com/10.1007/978-94-011-4455-1_1}.

\bibitem{takahashi_thermo_1996}
\bibinfo{author}{Takahashi, Y.} \& \bibinfo{author}{Umezawa, H.}
\newblock \bibinfo{title}{{THERMO} {FIELD} {DYNAMICS}}.
\newblock \emph{\bibinfo{journal}{International Journal of Modern Physics B}} \textbf{\bibinfo{volume}{10}}, \bibinfo{pages}{1755--1805} (\bibinfo{year}{1996}).
\newblock \urlprefix\url{https://www.worldscientific.com/doi/abs/10.1142/S0217979296000817}.

\bibitem{xu_quantum_2007}
\bibinfo{author}{Xu, X.-L.}, \bibinfo{author}{Li, H.-Q.} \& \bibinfo{author}{Wang, J.-S.}
\newblock \bibinfo{title}{Quantum fluctuations of mesoscopic {RLC} circuit involving complicated coupling in thermal squeezed state}.
\newblock \emph{\bibinfo{journal}{Physica B: Condensed Matter}} \textbf{\bibinfo{volume}{396}}, \bibinfo{pages}{199--206} (\bibinfo{year}{2007}).
\newblock \urlprefix\url{https://linkinghub.elsevier.com/retrieve/pii/S0921452607002256}.

\end{thebibliography}

\end{document}